\documentclass[reprint,prx,twocolumn,amsmath,amssymb]{revtex4-1}
\usepackage{mathrsfs}
\usepackage{mathtools}
\usepackage{xfrac}
\usepackage{color} 
\usepackage{pbox}

\usepackage{graphicx}
\usepackage{dcolumn}
\usepackage{bm}

\newcommand{\anni}[2]{\hat{\psi}_{#1}(#2)}
\newcommand{\crea}[2]{\hat{\psi}^{\dagger}_{#1}(#2)}

\newcommand{\ket}[1]{ | #1 \rangle }

\newcommand{\order}[1]{\langle #1 \rangle }

\newcommand{\vet}[1]{\vec{#1}}

\newcommand{\up}{\uparrow}
\newcommand{\down}{\downarrow}

\usepackage{hyperref}

\begin{document}

\title{Multi-$Q$ hexagonal spin density waves and dynamically generated spin-orbit coupling: \\ time-reversal invariant analog of the chiral spin density wave}

\author{J. W. F. Venderbos}
\affiliation{%
Department of Physics, Massachusetts Institute of Technology, Cambridge, Massachusetts 02139, USA
}%

\date{\today}

\begin{abstract}
We study hexagonal spin-channel (``triplet'') density waves with commensurate $M$-point propagation vectors.  We first show that the three $Q=M$ components of the singlet charge density and charge-current density waves can be mapped to multi-component $Q=0$ nonzero angular momentum order in three dimensions (3D) with cubic crystal symmetry. This one-to-one correspondence is exploited to define a symmetry classification for triplet $M$-point density waves using the standard classification of spin-orbit coupled electronic liquid crystal phases of a cubic crystal. 
Through this classification we naturally identify a set of noncoplanar spin density and spin-current density waves: the chiral spin density wave and its time-reversal invariant analog. These can be thought of as 3D $L=2$ and $4$ spin-orbit coupled isotropic $\beta$-phase orders. 
In contrast, uniaxial spin density waves are shown to correspond to $\alpha$ phases. The noncoplanar triple-$M$ spin-current density wave realizes a novel 2D semimetal state with three flavors of four-component spin-momentum locked Dirac cones, protected by a crystal symmetry akin to nonsymmorphic symmetry, and sits at the boundary between a trivial and topological insulator. In addition, we point out that a special class of classical spin states, defined as classical spin states respecting all lattice symmetries up to global spin rotation, are naturally obtained from the symmetry classification of electronic triplet density waves. These symmetric classical spin states are the classical long-range ordered limits of chiral spin liquids.  
\end{abstract}

\pacs{pacs}
\maketitle


\section{Introduction\label{sec:intro}}

Spin-orbit coupling, an intrinsic property of electrons in solids, is
at the root of many phenomena currently attracting a great deal of
attention in condensed matter physics. The topological insulators are a
prime example of a novel material class of which spin-orbit coupling
is a key characteristic~\cite{hasan10,qi11}. The importance of
spin-orbit coupling is most clearly reflected in the
spin-momentum locking of the celebrated Dirac surface states mandated by nontrivial bulk topology. In two dimensions graphene is a topological insulator in the presence of spin-orbit
coupling~\cite{kane05a}, however, in practice the spin-orbit
interaction turns out to be too weak. 

Whereas the topological insulators have been successfully
described in terms of non-interacting electron band theory, spin-orbit coupled Mott
insulators~\cite{jackeli09} are materials for which strong correlation effects are
important. The iridium-based oxides provide hallmark examples
of spin-orbit coupled materials where electron interactions have been proposed to give rise to intriguing phases of matter such as Kitaev spin
liquids~\cite{jackeli09,chaloupka10}, magnetic-order-induced Weyl semimetals~\cite{wan11},
and quantum critical nodal non-Fermi
liquids~\cite{savary14}. Spin-orbit coupling is key, since it
breaks spin-rotation symmetry and thus allows, for instance, for Kitaev-type terms in the spin
Hamiltonians of effective local moments describing these materials.

In both these cases, topological insulators and spin-orbit
coupled Mott insulators, the spin-orbit interaction is intrinsic and
of relativistic origin. Spin-orbit coupling can, however, also be
dynamically generated in nonrelativistic systems by
interactions. Instabilities of the Fermi liquid can lead to
condensation of particle-hole pairs into nematic phases with
anisotropic distortions of the Fermi surface~\cite{kivelson98,oganesyan01}, possibly in combination with time-reversal symmetry breaking~\cite{sun08}, or into isotropic
spin-orbit coupled phases~\cite{wu04}. Such electronic orders belong to the class of phases called electronic liquid crystals~\cite{kivelson98},  which are quantum analogs of classical liquid crystals, as they exhibit symmetry breaking but
remain metallic.

In this work, we show how aspects of these three phenomena involving spin-orbit coupling (i.e., dynamic generation of spin-orbit coupling, electron correlation, and nontrivial topology) come together in two-dimensional systems with hexagonal symmetry. Specifically, we study density wave formation, i.e., condensation of particle-hole pairs, at finite commensurate wave vectors associated with nested van Hove singularities. Our study comprises a classification of density waves in the spin channel, referred to as \emph{triplet} density waves~\cite{nayak00}, building on previous work which has focused on the singlet channel~\cite{venderbos15}. One of our main results is to show that multi-component density waves with nonzero (commensurate) wave vector in 2D can be mapped to and classified as condensates of particle-hole pairs with nonzero angular momentum in three dimensions (3D). For the triplet case this establishes a connection between the phenomenology of multi-component spin density wave states and spin-orbit coupled $\alpha$ and $\beta$ liquid crystal phases~\cite{wu04}. The latter are particle-hole analogs of the $A$ and $B$ phases of superfluid $^3$He~\cite{leggett75}.

Van Hove singularities connected by inequivalent commensurate wave vectors generically occur for electrons in simple hexagonal lattices such as the triangular and honeycomb lattices when doped to band structure saddle points.  Doped graphene is a notable example~\cite{neto09}. The van Hove singularities are located at the three centers of the Brillouin zone edges, i.e., the $M$ points (as shown in Fig.~\ref{fig:bzhexa}), and the wave vectors connecting them are the $M$-point vectors themselves.

A number of studies have addressed the effect of interactions between the three flavors of saddle point electrons, predicting exciting unconventional correlated phases such as topological chiral superconductivity, as well as Chern-insulating chiral spin density waves \cite{gonzalez08,nandkishore12a,kiesel12,yu12,wang12,kiesel13,wang13,nandkishore14,black14,jiang14,martin08,li12,makogon11,nandkishore12b,chern12}. These works highlight the rich physics expected in a broad class of (doped) hexagonal materials, with doped graphene as a concrete example. The purpose of this paper is to propose a comprehensive classification for density wave states that can arise when multi-flavor saddle-point electrons condense. Known phases such as the chiral and uniaxial spin density wave are shown to be natural products of such a classification. In addition, we find novel density wave states with topological quasiparticle spectra and hidden order.

We now give a brief overview of the content of this work and summarize the main results. 

\subsection*{Overview and main results}

In this work, we present a symmetry classification of hexagonal triplet
$M$-point order. At the heart of such classification is the notion of
extended point group symmetry. Extended point groups are crystal point groups supplemented
with those lattice translations that do not map the
enlarged unit cell, i.e., the unit cell defined by the
ordering vector of the density wave, to itself. Consequently, extended point groups
provide a natural and systematic way to study particle-hole
condensation at finite commensurate wave vector, as density waves can be classified in terms of
extended point group representations in the same way as angular
momentum channels are labeled by point group representations~\cite{sigrist91,platt13}. 

In previous work we have analyzed hexagonal
lattice $M$-point order in the singlet channel using extended point
group symmetry~\cite{venderbos15} and found a set of charge density ($s$) waves, in
addition to a set of time-reversal odd charge-current density ($d$) waves. Both
sets of orders correspond to nesting instabilities. On the basis of
that analysis, here we address the triplet variants of these
orders. The triangular and honeycomb lattices will serve as examples
of systems with hexagonal symmetry to which our analysis and results
apply, with an emphasis on the triangular lattice. 

A central theme of our study is dynamically generated spin-orbit coupling. 
By spin-orbit coupling here we mean the coupling of the spin and angular momentum of the particle-hole pairs in the condensed phase. The concept of dynamically generated spin-orbit coupling is introduced in more detail in the next section, where we consider $Q=0$ condensation. $Q=0$ condensation is caused by Pomeranchuk-type Fermi liquid instabilities and gives rise to phases that exhibit (anisotropic) Fermi surface distortions as a result of (rotational) symmetry breaking~\cite{kivelson98,oganesyan01,wu04,wu07,sun08}. In systems with hexagonal symmetry, spin-orbit coupled phases of the general form $ \order{\crea{\sigma}{\vet{k}}\anni{\sigma' }{\vet{k}}} =
\vet{\Delta}(\vet{k}) \cdot \vet{\sigma}_{\sigma\sigma'}$ with
$\vet{\Delta}(\vet{k})$ a linear combination of degenerate
$d$-wave form factors (relevant at van Hove doping), can be
distinguished by total angular momentum quantum numbers, as we explain in more detail in the following. One of such
orders, the $d+id$ $\beta$-phase~\cite{wu04,wu07}, is favored when nesting is
weak~\cite{maharaj13}.

The discussion of $Q=0$ $d$-wave orders, highlighting the coupling
of degenerate orbitals to spin, will set the stage for the
classification and study of triplet $M$-point order of the
general form $\order{\crea{\sigma}{\vet{k}+\vet{M}_\mu }\anni{\sigma'
  }{\vet{k}}} =  \vet{\Delta}_\mu(\vet{k}) \cdot
\vet{\sigma}_{\sigma\sigma'}$ ($\mu=1,2,3$, see Fig.~\ref{fig:bzhexa}). As a first step, we will consider the
nesting instabilities at the $M$ points in the spin channel, i.e., the
triplet $M$-point instabilities. Based on the nesting instabilities and their symmetry properties, we
will derive and discuss three main results.

\emph{(i)} We show that the $s$-wave and $d$-wave nesting instabilities map to sets of $L = 2$ and $4$ angular momenta, respectively, transforming as partners of representations of the cubic group $O_h$. This result is established by constructing a mapping between elements of the extended hexagonal point group $C'''_{6v}$ and elements of the cubic group, realizing an isomorphism between the two groups. Using this mapping, we will demonstrate that coupling the $L > 0$ orbitals to spin ($S$), in the same spirit as the $Q=0$ $d$-waves, defines a symmetry classification of hexagonal triplet $M$-point order: total angular momentum $J=L+S$ becomes a symmetry label for density waves. We argue that, from the perspective of symmetry and phenomenology, we can interpret distinct triplet $M$-point orders as electronic liquid crystal ($\alpha$ and $\beta$) phases of a 3D Fermi liquid with cubic symmetry. In addition, we present a dual interpretation in which electrons in a cubic crystal with intrinsic orbital degrees of freedom spontaneously develop orbital order. 

\emph{(ii)} Within the framework of the symmetry classification, we identify two distinct spin-orbit coupled cubic crystal $\beta$-phases, which are in correspondence with what we call scalar $M$-point density wave orders. They are scalar orders in the sense that they can be viewed as total angular momentum $J=0$ terms. The first originates from the set of charge density or $s$ waves and corresponds to a full spin-rotation symmetry broken spin density wave state, the so-called chiral spin density wave~\cite{martin08,li12,barros14,ghosh14}, associated with a gapped mean-field spectrum. The mean-field ground state is a Chern insulator. The second is a time-reversal invariant triplet $d$-wave or spin-current state, breaking no symmetries other than spin-rotation symmetry. We show that the mean-field ground state is a symmetry-protected $2D$ Dirac semimetal~\cite{young15}, protected by symmetries closely related to non-symmorphic crystal symmetry, and sits at the boundary between a trivial and a topological insulator. Using simple symmetry arguments, we discuss how symmetry breaking perturbations can drive the Dirac semimetal into either a trivial or topological electronic state. As such, the scalar (i.e., $J=0$) triplet $d$-wave state realizes a novel electronic phase, the dynamically generated 2D Dirac semimetal.

\emph{(iii)} The symmetry classification of triplet $M$-point order can be used to obtain the symmetric spin states of a given (hexagonal) lattice. Symmetric spin states are classical spin states that respect all symmetries of the crystal lattice up to a global rotation of all spins~\cite{messio11}. Such states are relevant in the context of magnetic materials, i.e., systems described by pure spin model Hamiltonians, as well as materials in which itinerant carriers are coupled to (large) localized spins. We will demonstrate that the symmetry classification provides a straightforward and constructive derivation of symmetric spin states. 

To summarize the organization of the paper, Sec.~\ref{sec:betaphase} will introduce dynamically generated spin-orbit coupling. In Sec.~\ref{sec:tripletm}, triplet $M$-point order is considered, by first discussing nesting instabilities and then proceeding to introduce the mapping from hexagonal to cubic symmetry. Using the mapping to define the symmetry classification, we identify and study two types of scalar triplet density wave states. Sec.~\ref{sec:lowenergy} aims at understanding the properties of the scalar triplet orders by focusing on low-energy electronic degrees of freedom. In Sec.~\ref{sec:classical}, we point out the connection to classical spin liquid states, and finally, in Sec.~\ref{sec:summary} we summarize and discuss the results presented.


\section{$Q=0$ triplet $d$-waves \label{sec:betaphase}}

In case of hexagonal $C_{6v}$ symmetry, the $d$-wave channel is two-fold degenerate, with the $d$-wave orbitals $(d_{x^2-y^2},d_{xy})$ transforming as partners of a two-dimensional representation. Therefore, if the leading instability is in the $Q=0$  $d$-wave channel, a general linear combination of the two $d$-wave orbitals is allowed. This situation has been shown to occur for the triangular and honeycomb lattices electrons at van Hove doping, when nesting is weak~\cite{maharaj13,valenzuela08}. In the singlet channel only real superpositions of the two degenerate orbitals are allowed. In the triplet channel, however, both real and complex or ``chiral'' linear combinations are possible.  

In the triplet channel, real and chiral linear combinations of the $d$ waves are lattice analogs of the $\alpha$ and $\beta$ electronic liquid crystal phases with dynamically generated ``spin-orbit coupling'' in continuum Fermi liquids~\cite{wu04,wu07}. Both in the $\alpha$ and $\beta$ phases spin-rotation symmetry is spontaneously broken. In the $\alpha$ phase the spin order is uniaxial and spin rotation is only partially broken. Spatial rotations are broken due to $d$-wave nature of the orbital angular momentum. In contrast, spin-rotation symmetry is fully broken in the $\beta$ phase, yet the $\beta$ phase is isotropic due to the coupling of spin and orbital angular momentum: combined spin and spatial rotations leave the state invariant. The isotropy can be thought of as a consequence of two angular momenta adding to form a rotationally invariant singlet state. 

Let us show this more explicitly. We collect the two $d$-wave orbitals in a vector $\vet{\lambda}(\vet{k}) =
(\lambda_{d_{1}} ,\lambda_{d_{2}} )$ (explicit expressions of lattice form factors can be found in Table~\ref{tab:trifunctions} of Appendix~\ref{app:gt}) and then write the $\alpha$ phase as
\begin{gather} \label{eq:alphatri}
\order{\crea{\sigma }{\vet{k}}\anni{\sigma'}{\vet{k}}} =
\vet{\Delta}_\alpha \cdot \vet{\lambda}(\vet{k}) \sigma^3_{\sigma\sigma'}.
\end{gather}
The two-component order parameter $\vet{\Delta}_\alpha$ is real, which follows from the requirement of Hermiticity. As a result, $\vet{\Delta}_\alpha$ is a nematic order parameter breaking lattice rotational symmetry~\cite{wu07}. Due to the uniaxial spin polarization along the $z$ axis, spin-rotation symmetry is only partially broken.  

Instead, the $\beta$ phase takes the form of a dot product of $d$ waves and spin, $\vet{\lambda} \cdot \vet{\sigma}$, given by
\begin{gather} \label{eq:betatri}
\order{\crea{\sigma }{\vet{k}}\anni{\sigma'}{\vet{k}}} =  \Delta_\beta[\lambda_{d_{1}}(\vet{k}) \sigma^1 _{\sigma\sigma'}+\lambda_{d_{2}}(\vet{k}) \sigma^2 _{\sigma\sigma'}].
\end{gather}
This is a chiral superposition of $d$-waves, as is easily seen by considering the off-diagonal elements $\order{\crea{\down }{\vet{k}}\anni{\up}{\vet{k}}} = \Delta_\beta (\lambda_{d_{1}}+i \lambda_{d_{2}}) \sim d+id $. As a  result of the spin-orbit coupling $\vet{\lambda} \cdot \vet{\sigma}$ momentum and spin are locked together in such a way that (properly chosen) simultaneous rotations make the $\beta$ phase isotropic. 

As a spoiler for the next section, the coupling of $d$-waves and spin in the $\beta$-phase can be obtained more formally by the standard recipe for addition of angular momenta in a crystal. In hexagonal symmetry the $d$-waves transform as the nematic doublet $E_2$, and the spin components $(\sigma^1,\sigma^2)$ as the vector doublet $E_1$. Addition of angular momenta is then given by $E_2 \times E_1 = B_1 + B_2 + E_1$. The first two terms are scalars, $\vet{\lambda} \cdot \vet{\sigma}$ and $\hat{z}\cdot \vet{\lambda} \times \vet{\sigma}$, which both correspond to isotropic $\beta$-phases.

It should be noted that even though the $\alpha$ phase is non-chiral whereas the $\beta$ phase is chiral, both break time-reversal symmetry since both are triplet $d$-wave states. In addition, it should be noted that unitary rotations in spin space will yield equivalent states in case of both types of phases, since the normal state is spin-rotation invariant. 

When the dominant instability is in the $Q=0$ $d$-wave particle-hole channel, it was shown that the chiral $d+id$ $\beta$-phase is favored~\cite{maharaj13}. This result mirrors the result found in the Cooper channel, in which case chiral $d+id$ superconductivity is  favored~\cite{nandkishore12a,kiesel12}. These findings are rooted in hexagonal symmetry and therefore apply to both the triangular and honeycomb lattices. An expression similar to Eq.~\eqref{eq:betatri}, taking the sublattice structure into account, can be written for the honeycomb lattice.

We summarize this section by noting that hexagonal $Q=0$ triplet states are examples of dynamically generated spin-orbit coupling. In particular, the $\beta$-phase of Eq.~\eqref{eq:betatri} is a spin-orbit coupled scalar, analogous to total angular momentum $J= L+S=0$ states arising from addition of two angular momenta $L$ and $S$.


\section{Hexagonal triplet $Q=M$ states\label{sec:tripletm}}

Particle-hole condensation at finite wave vector is expected when the Fermi surface is strongly or perfectly nested by that wave vector. For hexagonal lattices such as the triangular and honeycomb lattices, Fermi surface nesting can occur at three inequivalent wave vectors $\vet{M}_\mu$. In this section, we study triplet density waves of hexagonal lattices with commensurate ordering vectors $\vet{M}_\mu$ having the general form
\begin{gather}
\order{\crea{\sigma}{\vet{k}+\vet{M}_\mu }\anni{\sigma' }{\vet{k}}} =  \vet{\Delta}_\mu(\vet{k}) \cdot \vet{\sigma}_{\sigma\sigma'}, 
\end{gather}
where $\vet{\Delta}_\mu(\vet{k}) \cdot \vet{\sigma}_{\sigma\sigma'}$ is the triplet version of the singlet term $\Delta_\mu(\vet{k})\delta_{\sigma\sigma'}$ (suppressing sublattice indices). As a first step towards this goal, we derive all distinct nesting instabilities using a simple algebraic approach. We establish the symmetry of the nesting instabilities using extended point group representations. Based on that we define a symmetry classification of triplet nesting instabilities in terms of spin-orbit coupling, in close analogy with the spin-orbit coupled $Q=0$ phases. To this end, we introduce a mapping from hexagonal symmetry to cubic symmetry. We then analyze a specific class of spin-orbit coupled orders, the total angular momentum $J=0$ orders, in more detail.

\subsection{Nesting instabilities\label{ssec:nesting}}

\begin{figure}
\includegraphics[width=\columnwidth]{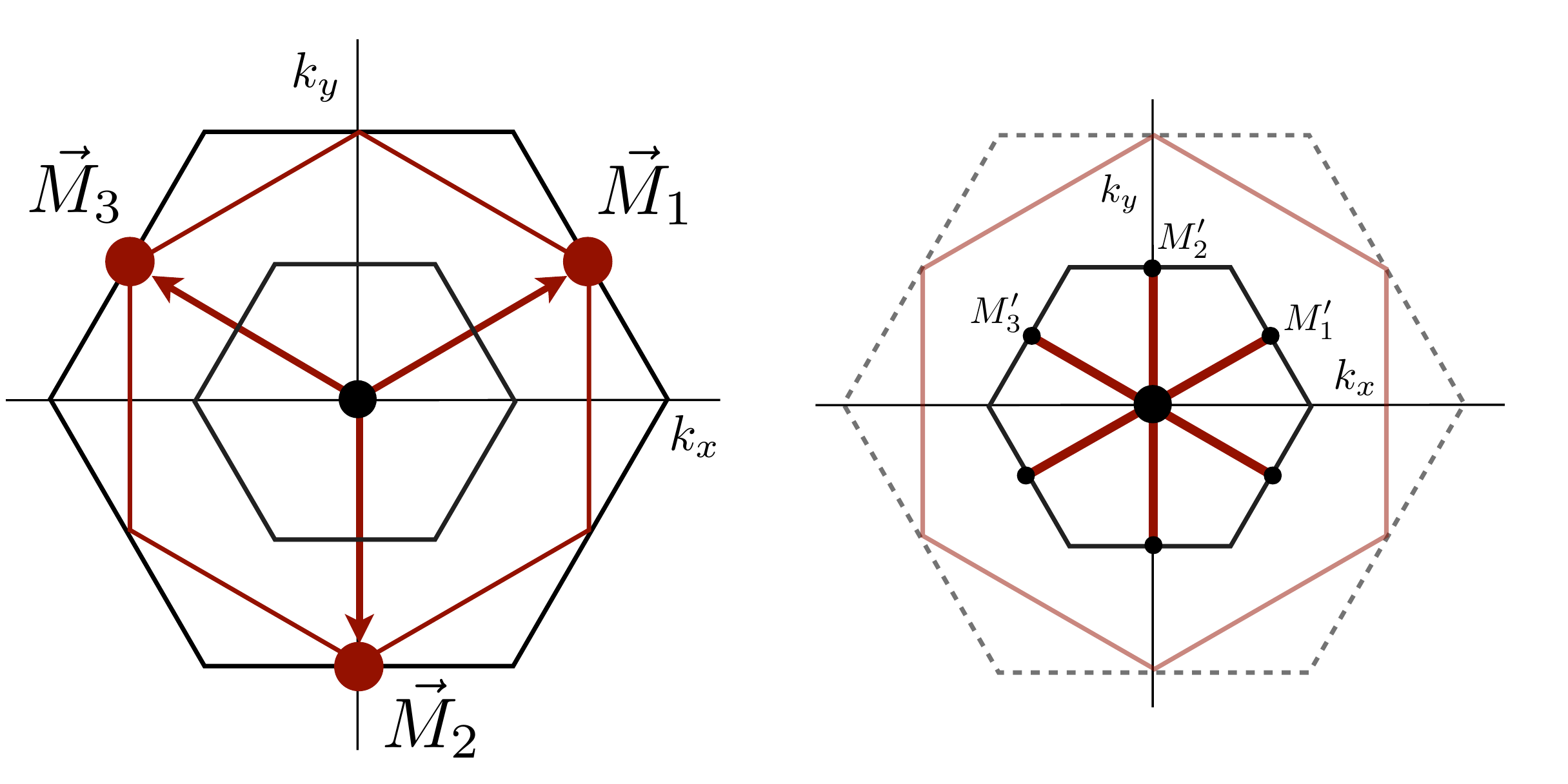}
\caption{\label{fig:bzhexa} (Left) Brillouin zone (BZ) of hexagonal lattices with special $M$-point momenta $\vet{M}_\mu$. Red hexagon represents the Fermi surface at Van Hove filling; small black hexagon is the folded BZ of the quadrupled unit cell. (Right) Folded BZ and high symmetry $M'$-points; red line segments are the folded Fermi surface. Dashed lines are the unfolded BZ (black) and Fermi surface (red).}
\end{figure}

The analysis of hexagonal lattice nesting instabilities in the presence of spin is a straightforward extension of the analysis without spin degree of freedom~\cite{venderbos15}. The goal is to construct an algebra of the van Hove electrons and use that algebra to determine the nesting instabilities. For lattices with hexagonal symmetry, the three van Hove singularities are located at the inequivalent $M$-point momenta $\vet{M}_\mu$, shown in Fig.~\ref{fig:bzhexa}. The algebra of the $M$-point electrons can be expressed in terms of the van Hove electron operators $\hat{\Phi}$,
\begin{gather} \label{eq:vanhovehexa}
\hat{\Phi} =  \begin{pmatrix} \anni{\sigma}{\vet{M}_1} \\ \anni{\sigma}{\vet{M}_2}   \\ \anni{\sigma}{\vet{M}_3}   \end{pmatrix}  \equiv \begin{pmatrix}\hat{\psi}_{1\sigma } \\ \hat{\psi}_{2\sigma }    \\ \hat{\psi}_{3\sigma }   \end{pmatrix},
\end{gather}
where $\sigma$ labels the spin. The full hexagonal van Hove algebra is then defined by the bilinears $\hat{\Phi}^\dagger \Lambda^i\sigma^j \hat{\Phi}$, where $\Lambda^i$ is the set of Gell-Mann matrices, i.e., the generators of SU(3) (we also include the identity matrix), and $\sigma^j$ are Pauli matrices [generators of SU(2)] which act on the electron spin. The matrices $\Lambda^i$ act on the species index $\mu$ corresponding to $\vet{M}_\mu$. The explicit form of the set of eight $\Lambda^i$ is given in Appendix~\ref{app:malgebra}. We find it convenient to group them into three subsets, denoted by $\vet{\Lambda}_a$, $\vet{\Lambda}_b$, and $\vet{\Lambda}_c $. 

From the set of SU(3) generators we can form three SU(2) sub-algebras, each of which acts in the subspace of a pair of van Hove electrons. In this way the sub-algebras are associated with the three ways of connecting two $M$-points, i.e., coupling the van Hove electrons. For instance, the pair of van Hove electrons $ \hat{\psi}_1$ and $ \hat{\psi}_2$ is connected by wave vector $\vet{M}_3$. The SU(2) algebra specified by $(\Lambda^1_a,\Lambda^1_b,\Lambda^1_c)$ acts in the subspace $(  \hat{\psi}_1 ,\hat{\psi}_2  )$ and governs the nesting instabilities at wave vector $\vet{M}_3$. The matrix $\Lambda^1_c$ is diagonal and describes the population imbalance of van Hove electrons $ \hat{\psi}_1 $ and $\hat{\psi}_2 $. All bilinears that do not commute with $\hat{\Phi}^\dagger \Lambda^1_c\hat{\Phi}$ give rise to nesting instabilities. The non-commuting set of matrices is given by $\Lambda^1_a$ and $\Lambda^1_a\sigma^j$, corresponding to charge and spin density waves, respectively, in addition to $\Lambda^1_b$ and $\Lambda^1_b\sigma^j$, which correspond to charge currents and spin currents. 

Nesting instabilities at the wave vectors $\vet{M}_1$ and $\vet{M}_2$ are obtained by either explicitly constructing the van Hove sub-algebras of the doublets $(\hat{\psi}_2 ,\hat{\psi}_3)$ and $(\hat{\psi}_3 ,\hat{\psi}_1)$, or more directly by using rotational symmetry. One finds that the bilinears specified by the matrices $\vet{\Lambda}_a$ correspond to three degenerate charge density wave instabilities at the three different wave vectors, whereas the set $\vet{\Lambda}_b$ corresponds to three degenerate charge-current density waves. In the spin channel, the set $\vet{\Lambda}_a\sigma^j$ describes spin density waves and the set $\vet{\Lambda}_b\sigma^j$ describes spin-current density waves. 


Let us consider the symmetry properties of the nesting instabilities. Charge density wave order given by $\vet{\Lambda}_a$ respects time-reversal symmetry while orbital current order $\vet{\Lambda}_b$ is odd under time reversal. Since spin is odd under timereversal, the triplet orders $\vet{\Lambda}_a\sigma^j$ (spin density waves) and $\vet{\Lambda}_b\sigma^j$ (spin-current density waves) are odd and even under time reversal, respectively. As far as spatial symmetries are concerned, the components of $\vet{\Lambda}_a$ transform as the representation $F_1$ of the extended point group $C'''_{6v}$, and the components of $\vet{\Lambda}_b$ as $F_2$ (see also Appendix~\ref{app:gt}). The $F_1$ representation describes $s$-wave form factors at each wave vector and the $F_2$ representation describes $d$-wave form factors, so we may use these symmetry labels to refer to charge and orbital current order (see also Table~\ref{tab:correspond}). In particular, we refer to $\vet{\Lambda}_a\sigma^j$ as triplet $s$-wave or triplet $F_1$ order, and similarly for $\vet{\Lambda}_b\sigma^j$. We assume absence of spin-orbit coupling in the normal state leading to a full SU(2) rotation symmetry in spin space. Therefore, we simply distinguish singlet and triplet instabilities. We thus obtain
\begin{gather}
F_1 \; \rightarrow \; \vet{\Lambda}_a\; (+), \; \vet{\Lambda}_a\sigma^j \; (-),  \nonumber \\
F_2 \; \rightarrow \; \vet{\Lambda}_b \; (-), \; \vet{\Lambda}_b\sigma^j \; (+), \label{eq:nestsym}
\end{gather}
where $(\pm)$ indicates even or odd under time-reversal.

\subsection{Global spin rotation and mapping to cubic $L > 0$ angular momentum phases\label{ssec:soc}}

The purpose of this section is to establish the correspondence between symmetries of the density waves in 2D and symmetries of angular momenta in 3D. This requires studying the action of symmetries of the hexagonal lattice on the density waves in more detail. To proceed, we therefore explore the properties of the $M$-point representation of hexagonal symmetry. The $M$-point representation is simply given by the action of symmetries on the three $M$ points (or, equivalently, the three density wave components). It can be defined in terms of a vector $\vet{v}$, the components of which are the linearly independent functions describing modulations with $M$-point propagation vectors. It is given by 
\begin{gather}
\vet{v}(\vet{x}) = \begin{pmatrix} \cos \vet{M}_1\cdot\vet{x} \\ \cos \vet{M}_2\cdot\vet{x}  \\  \cos \vet{M}_3\cdot\vet{x}  \end{pmatrix},
\end{gather} 
where $\vet{x}$ labels the sites of the triangular Bravais lattice. The effect of elementary translations $T(\vet{a}_i)$ ($i=1,2,3$) on $\vet{v}$ is given by $\vet{v}(\vet{x}+\vet{a}_i) = G_i \vet{v}(\vet{x})$. Here, $\vec{a}_{1,2} = (1,\pm \sqrt{2})/2$ and $\vec{a}_{3} =(-1,0)$. We define the action of the group generators $C_6$ and $\sigma_v$ on $\vet{v}$ as $X\vet{v}(\vet{x})$ and $Y\vet{v}(\vet{x})$, respectively, which are given by 
\begin{gather} 
X = \begin{pmatrix}  0 & 1 & 0 \\ 0 & 0 & 1 \\1 & 0 & 0 \end{pmatrix}, \quad Y = \begin{pmatrix}  0 & 0 & 1 \\ 0 & 1 & 0 \\ 1 & 0 & 0 \end{pmatrix},
\end{gather}
(more details in Appendix~\ref{app:malgebra}). All matrices $G_i$ are diagonal, an example is $G_1 = \text{diag}(-1,-1,1)$. Consequently, the $O(3)$ matrices $G_i$, $X$, and $Y$ generate a representation of the extended hexagonal point group $C'''_{6v}$.

A general modulation of electron density with $M$-point propagation vectors can be written as a linear combination of the three components of $\vec v$, $\vet{w}\cdot \vet{v}(\vet{x})$, where $\vet{w}$ is an arbitrary vector. Except for certain special cases, such linear combinations will not respect lattice symmetries. 
This is certainly true for the translations: $G_i \vet{w} \neq \vet{w}$ for at least one $G_i$ and general $\vet{w}$. When spin is taken into account, however, a more careful consideration of the symmetries of $M$-point order is required. We now demonstrate this based on the simplest case of $M$-point modulations described by single $\vet{w}$ [i.e., $\vet{w}\cdot \vet{v}(\vet{x})$], hence avoiding unessential details such as sublattice structure. 

General modulations of spin density, with spin represented by the Pauli matrices $\vet{\sigma}$, are written in terms of a matrix $W$ as
\begin{gather} \label{eq:spinorderm}
\vet{\sigma} \cdot   W \cdot \vet{v}(\vet{x}).
\end{gather}
This can be thought of as a vector $\vet{w}$ for each spin direction. The effect of a translation is then expressed as $\vet{\sigma} \cdot   W \cdot G_i\vet{v}(\vet{x})$, where the $G_i$ act on $W$ from the right. Note that matrices acting from the right act on the $M$-point indices $\mu$, whereas matrices acting from the left of $W$ act on the spin indices $i$ (i.e., $\vet{\sigma} \cdot   W \cdot \vet{v} = \sigma^iW_{i\mu}v_\mu$). Due to the spin degree of freedom, it can occur that in some cases multiplication by $G_i$ from the right is equal to multiplication by some $O(3)$ matrix $\mathcal{R}_i$ from the left, 
\begin{gather} 
\vet{\sigma} \cdot   W \cdot G_i\vet{v}  = \vet{\sigma} \cdot  \mathcal{R}_i W \cdot \vet{v}.
\end{gather}
In this way, the action of translations is carried over to spin space, since $\mathcal{R}_i$ is a global spin-rotation matrix. A trivial example of this is the identity matrix, i.e. $W=I$, in which case $\mathcal{R}_i=G_i$~\cite{note3}. 

The key implication of the relation $WG_i = \mathcal{R}_i W$ is the equivalence of translations and spin rotations. Specifically, it implies that the term $\vet{\sigma} \cdot   W \cdot \vet{v} = \vet{\sigma} \cdot   \vet{W} $ (with $\vet{W}  \equiv W \cdot \vet{v}$) can be made invariant under translations with the help of a unitary matrix $U \in $ SU(2) associated with $\mathcal{R}$, expressed as
\begin{gather} 
\vet{\sigma} \cdot \vet{ W}  = U^\dagger \vet{\sigma} \cdot ( \mathcal{R}\vet{ W} ) U. 
\end{gather}
The unitary matrix $U$ acts on the matrix components of the $\sigma^j$, and through the correspondence with $\mathcal{R}$ implements the global spin rotation. Hence, even though $M$-point modulations would seem to break translational symmetry, in certain cases (for certain $W$) they are invariant up to global spin rotation~\cite{martin08}. 

The significance of these global spin rotations is the resulting invariance of the Hamiltonian $H(\vet{k})$. A translation combined with a global spin rotation leaves the Hamiltonian invariant and this effective symmetry will lead to degeneracies of the spectrum. 

The same argument applies to the point group generators $C_6$ and $\sigma_v$, in which case the global spin rotation matrices are denoted as $\mathcal{R}_X$ and $\mathcal{R}_Y$. It is important, however, to distinguish proper and improper $\mathcal{R}$, since one can only associate an SU(2) matrix to a proper $\mathcal{R}$. Improper rotations acquire an extra minus sign, i.e., $\vet{\sigma} \cdot ( \mathcal{R}\vet{ W} ) = - U \vet{\sigma} \cdot \vet{ W} U^\dagger$, as a consequence of inversion. Again taking the simplest case $W=I$ as an example, it is easy to see that all reflections, which are composed of $Y$, are improper. As a result, $M$-point spin density modulations of the form $\vet{\sigma}\cdot  \vet{ v}(\vet{x}) $ are odd under reflections. 

These considerations demonstrate that in order to properly analyze the symmetry properties of triplet $M$-point density waves, it is important to take the global spin rotations into account. 

The expression $\vet{\sigma} \cdot   W \cdot \vet{v} = \vet{\sigma} \cdot   \vet{W} $ in~\eqref{eq:spinorderm} bears a suggestive resemblance to the familiar spin-orbit coupling term $\vet{S}\cdot \vet{L}$, where $\vet{S}$ is spin and $\vet{L}$ is orbital angular momentum. Indeed, as is clear from the preceding discussion, $\vet{\sigma} \cdot   \vet{W} $ implies an entangling of spatial symmetries and spin. We now show how this resemblance can be formalized by exploiting the mapping between the extended hexagonal point group $C'''_{6v}$ and the cubic group $O_h$. 
We then explain how such mapping provides a way to classify and determine the symmetry of triplet $M$-point density wave orders. 

We start by explicitly constructing the mapping from the group $C'''_{6v}$ to the cubic group $O_h$. It is easy to see that the matrices $G_i$, $X$, and $Y$, associated with the generators of $C'''_{6v}$, are $O(3)$ rotation matrices. The key observation is that they correspond to rotations that leave a cube invariant. For instance, the matrices $G_i$ are two-fold rotation about the principal axes, and $X$ is a three-fold rotation about the body diagonal. As a result, the $M$-point representation also realizes a representation of $O_h$. This is, however, not a faithful representation, since the two-fold rotation $C_2 = C_6^3$ in $C'''_{6v}$ is represented by the identity, i.e., $X^3=1$. A faithful representation is obtained by redefining the generator matrix of $C_6$ as $-X$. In this way, the twofold rotation in $C'''_{6v}$ is mapped to the inversion in $O_h$: $(-X)^3=-1$. This establishes a one-to-one correspondence between elements of $C'''_{6v}$ and $O_h$. In particular, the representations of the two groups and their basis functions are in one-to-one correspondence. For instance, the cubic representation generated by $\{ G_i, -X, Y \}$ is given by $T_{1u}$ symmetry, and it is simple to check that the representation generated by $\{ G_i, X, Y \}$ is $T_{2g}$.   The three-component nature of representations of $C'''_{6v}$ is rooted in nonzero ($M$-point) linear momenta, whereas the three-component nature of $O_h$ representations is due to nonzero \emph{angular} momenta. Angular momentum orbitals with $T_{1u}$ and $T_{2g}$ symmetry are given by $(k_x,k_y,k_z)$ and $(k_yk_z,k_zk_x,k_xk_y)$. 

From this follows one of the main results of this paper: hexagonal density waves with $M$-point wave vectors can be mapped to $Q=0$ nonzero angular momentum ($L \neq 0$) order in a three-dimensional cubic crystal. Another way of stating it is that particle-hole pairs with finite momentum in 2D can be uniquely associated with particle-hole pairs with nonzero angular momentum in 3D. The latter belong to the class of liquid crystal phases. What is the implication of this correspondence? To answer that question, we first relate the $C'''_{6v}$ representations $F_1$ and $F_2$, which describe the symmetry of the charge and charge-current density waves, to representations of the cubic group. The former corresponds to $T_{2g}$, whereas the latter is generated by $\{ G_i, X, -Y \}$ and corresponds to $T_{1g}$. The cubic representations $T_{2g}$ and $T_{1g}$ describe three-fold degenerate orbitals coming from $L=2$ and $4$ angular momenta, respectively. Evidently, the charge density $s$ waves, which have $F_1$ symmetry, are mapped to the $T_{2g}$ orbitals $(k_yk_z,k_zk_x,k_xk_y)$. The charge-current $d$-density waves with $F_2$ symmetry are mapped to $T_{1g}$ orbitals, which are listed in Table~\ref{tab:correspond}.

\begin{table}[t]
\centering
\begin{ruledtabular}
\begin{tabular}{ccccc}
Order type & \multicolumn{2}{c}{Extended group $C'''_{6v}$}  & \multicolumn{2}{c}{Cubic group $O_h$}   \\ 
\hline
\pbox{4cm}{Charge \\ ($s$-wave)} & $F_1$  & $\begin{pmatrix} 1 \\1 \\1 \end{pmatrix}$ & $T_{2g}$ & $\begin{pmatrix} k_yk_z \\ k_zk_x \\ k_xk_y \end{pmatrix}$ \\
\pbox{4cm}{Charge-current \\ ($d$-wave)}& $F_2$  & $i\begin{pmatrix} k^2_3-k^2_1 \\ k^2_1-k^2_2 \\k^2_2-k^2_3 \end{pmatrix} $ & $T_{1g}$ &  $\begin{pmatrix}  k_yk_z(k^2_y-k^2_z) \\k_zk_x(k^2_z-k^2_x) \\ k_xk_y(k^2_x-k^2_y \end{pmatrix}$\\
\end{tabular}
\end{ruledtabular}
 \caption{Table showing the correspondence of representations of the extended hexagonal point group $C'''_{6v}$ and the cubic group $O_h$. In addition, we list the angular momentum functions transforming as these representations. Hexagonal $F_1$ and $F_2$ order is shown to be of $s$- and $d$-wave type, respectively. Their cubic equivalents are composed of $L =2$ ($d$ wave) and $L =4$ orbitals. Note that hexagonal $d$-wave $M$-point order (i.e., $F_2$) is imaginary and breaks time-reversal symmetry. Here, $k_{1,2,3} = \vet{k}\cdot \vet{a}_{1,2,3} $, where $\vet{a}_{1,2,3} $ are the lattice vectors of the triangular Bravais lattice. }
\label{tab:correspond}
\end{table}

The usefulness of the mapping between hexagonal and cubic symmetry becomes apparent once spin is considered. It can be stated as follows: spin-orbit coupled cubic liquid crystals, i.e., spin-rotation symmetry broken $\alpha$ and $\beta$ phases formed from $T_{1g}$ and $T_{2g}$ orbitals, can be used to define a classification of hexagonal triplet density wave states. Since each spin-orbit coupled $\alpha$ and $\beta$ phase uniquely corresponds to a density wave state, a classification of the former implies a classification of the latter. This is a second key result of this paper and we now demonstrate this in detail. 

The spin-orbit coupled cubic liquid crystal phases with $T_{1g}$ and $T_{2g}$ angular momenta are direct analogs of the $\alpha$- and $\beta$-phases discussed in the previous section. Consider first the $\beta$ phase case. Spin angular momentum $\vec{\sigma}$ transforms as $T_{1g}$ under cubic symmetry. The $\beta$ phase is a spin-orbit coupled state for which only total angular momentum is a good quantum number. Taking the $T_{2g}$ orbitals as a first example, the good quantum numbers in a cubic crystal are obtained from the product representation $T_{1g} \times T_{2g}$ which is decomposed as $A_{2g} + E_g + T_{1g} + T_{2g}$. This is the lattice analog of the addition of a pair of angular momenta $L=1$ (orbital) and $S=1$ (spin) in the presence of full rotational symmetry, giving total angular momentum $J= L+S=0,1,2$. The term $A_{2g}$ should be interpreted as the $J=0$ case and corresponds to an isotropic $\beta$-phase. Collecting the $T_{2g}$ orbitals in a vector $\vet{d}(\vet{k}) = (k_yk_z,k_zk_x,k_xk_y)$, and denoting electron operators of a 3D Fermi liquid (with cubic crystal anisotropy) by $\hat{\chi}(\vet{k})$, the spin-orbit coupled $\beta$ phase takes the form 
\begin{gather} \label{eq:cubicbeta}
\order{\hat{\chi}^\dagger_{\sigma}(\vet{k})\hat{\chi}_{\sigma' }(\vet{k})} =  \Delta_{\beta} \; \vet{d}(\vet{k}) \cdot \vet{\sigma}_{\sigma\sigma'}, 
\end{gather}
which is a 3D analog of Eq.~\eqref{eq:betatri}. Other terms in the decomposition of $T_{1g} \times T_{2g}$ correspond to multi-component anisotropic spin-orbit coupled liquid crystals. 

The terms in the decomposition are symmetry labels (quantum numbers) for spin-orbit coupled liquid crystals and provide a way to classify these phases, in the same way as $J=0,1,2$ classifies total angular momentum. Then, as a consequence of the one-to-one correspondence between the cubic orbitals and hexagonal $M$-point density waves, this implies a symmetry classification of the triplet $M$-point density waves. Each spin-orbit coupled liquid crystal state can be mapped back to a unique density wave state, and its symmetry follows directly from the one-to-one correspondence. A key property of the classification obtained in this way is that it manifestly takes global spin-rotation invariance into account.  

To illustrate this, let us consider the $T_{2g}$ $\beta$-phase with $A_{2g}$ symmetry. Comparing the character tables of $O_h$ and $C'''_{6v}$ we find that $A_{2g}$ symmetry in the former corresponds to $A_2$ symmetry in the latter. Importantly, the $A_2$ representation is a translationally invariant representation. This shows that global spin-rotation equivalence is manifest, since the only way to preserve translations for $M$-point order is to combine them with global spin rotations. Note that a spin density wave with $A_2$ symmetry breaks reflection symmetry. 

Similar to the case of $T_{2g}$ orbitals, we can obtain the $\beta$-phase in case of the $T_{1g}$ orbitals. Spin-orbit coupling leads to the decomposition
\begin{gather}
T_{1g} \times T_{1g}= A_{1g} + E_g + T_{1g} + T_{2g}. \label{eq:cubicorder}
\end{gather}
Here the $A_{1g}$ term corresponds to a $J=0$ $\beta$-phase. $A_{1g}$ symmetry implies full invariance under cubic symmetry and therefore full invariance under hexagonal symmetry. As a result, the triplet spin-current $d$-density wave which uniquely corresponds to the $A_{1g}$ $\beta$-phase breaks no spatial symmetries.

In addition to the $\beta$ phase we can also consider the analog of the $\alpha$ phase. For instance, the cubic $\alpha$ phase with constituent $d$ wave orbitals $\vet{d}(\vet{k}) = (k_yk_z,k_zk_x,k_xk_y)$ is written as
\begin{gather} \label{eq:cubicalpha}
\order{\hat{\chi}^\dagger_{\sigma}(\vet{k})\hat{\chi}_{\sigma' }(\vet{k})} =  \vet{\Delta}_{\alpha} \cdot \vet{d}(\vet{k})  \sigma^j_{\sigma\sigma'},
\end{gather}
which should be compared to Eq.~\eqref{eq:alphatri}. Here, $j$ is a given direction in spin space, for instance, the global $z$ axis, making it a uniaxial spin ordered state with only partially broken spin-rotational symmetry. We will see in the following that the $\alpha$ phase corresponds to a uniaxial spin density wave. 

Let us summarize the main result of this section. By establishing a one-to-one correspondence between the cubic group $O_h$ and the hexagonal group $C'''_{6v}$, we have reformulated hexagonal $M$-point order associated with nesting instabilities in terms of electronic liquid crystal states with nonzero angular momentum in a cubic crystal. Coupling spin and orbital angular momentum allowed us to assign a symmetry label to hexagonal triplet ordering in a way that gives the correct symmetry of the orders. 

Specifically, we obtain two special triplet density wave states (``$\beta$ phases'') labeled by nondegenerate representations and transforming as scalars. These can be viewed as isotropic total angular momentum $J=0$ states. Due to this, in what follows we will refer to these orders as \emph{scalar} triplet states, or alternatively a scalar spin-orbit coupled states. In the remainder of this section we focus on these two scalar triplet states with high symmetry and study them in more detail based on their realization on the triangular and honeycomb lattices.

\subsection{$s$-wave triplet states: Spin density waves\label{ssec:swave}}

\begin{figure}
\includegraphics[width=0.9\columnwidth]{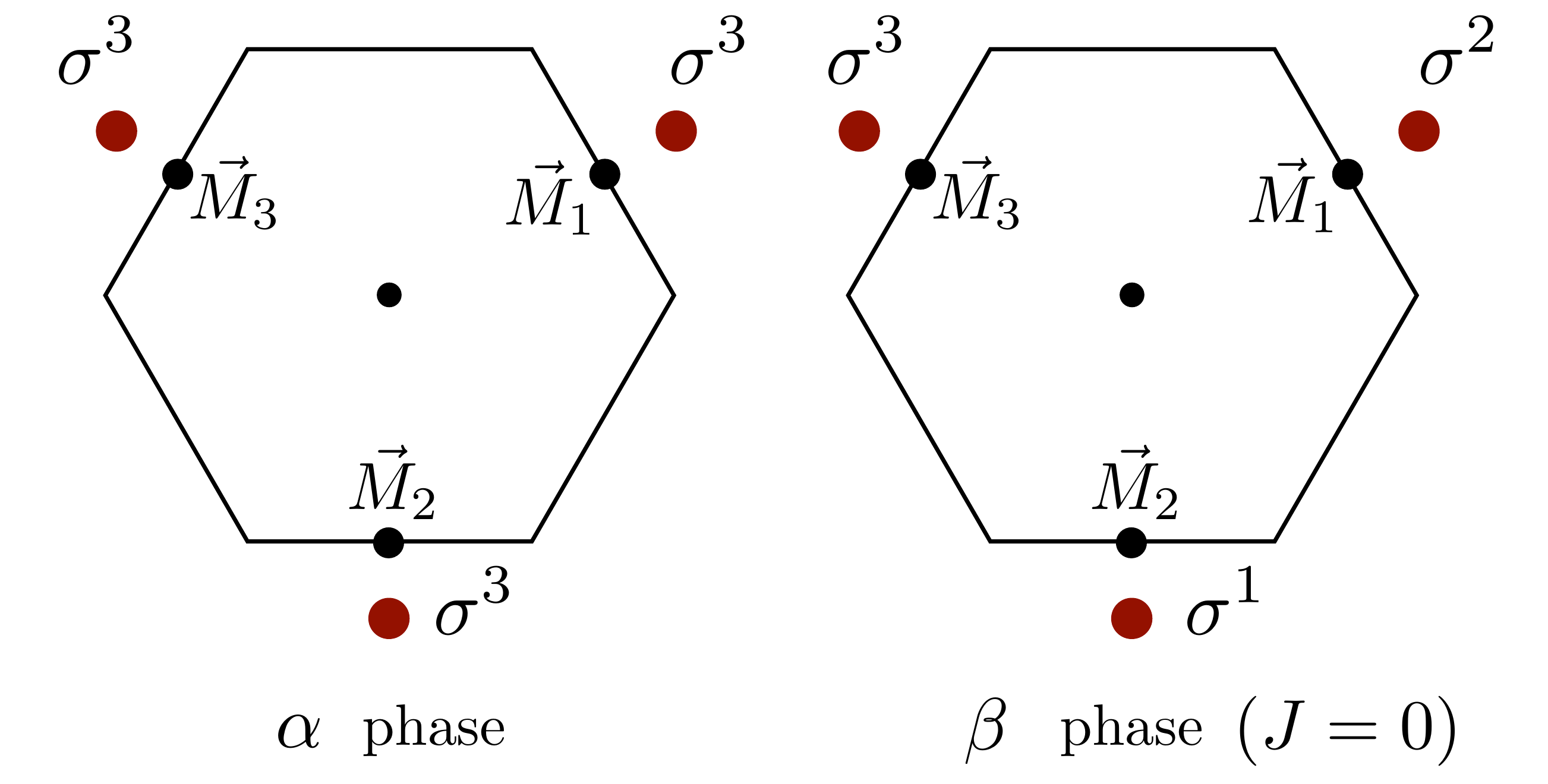}
\caption{\label{fig:f1spin} Schematic representation of $F_1$ $s$-wave triple-$M$ order. (Left) Each order parameter component (i.e., $M$-point) is associated with an $s$-wave orbital and the same spin matrix $\sigma^3$ giving a uniaxial state. This state is related to the cubic $\alpha$ phase of Eq.~\eqref{eq:cubicalpha}. (Right) Same as on the left, except that each $M$-point is associated with a different spin matrix, resulting in the chiral state of~\eqref{eq:trichiral} and \eqref{eq:honchiral}. The chiral states are related to the cubic $\beta$-phase of Eq.~\eqref{eq:cubicbeta}. }
\end{figure}

Triplet $s$-wave states are the spin density waves derived from the $F_1$ representation and correspond to the nesting instabilities given by $\vet{\Lambda}_a\sigma^j$ in Eq.~\eqref{eq:nestsym}. The symmetry classification of triplet $s$ waves gives rise to a scalar triplet state with $A_2$ symmetry, which is the $T_{2g}$ $\beta$ phase mapped back to a density wave state. Here, we construct this scalar order explicitly for both the triangular and honeycomb lattices and study its properties. In both cases, the scalar order is obtained by pairing each $M$-point component $\vet{M}_\mu$ with a spin component $\sigma^j$ (i.e., Pauli matrix). This is schematically shown in Fig.~\ref{fig:f1spin}. For the triangular lattice, this directly leads to
\begin{gather} \label{eq:trichiral}
\order{\crea{\sigma}{\vet{k}+\vet{M}_\mu }\anni{\sigma' }{\vet{k}}} =  \Delta_{A_2} \;  \sigma^\mu_{\sigma\sigma'}. 
\end{gather}
In case of the honeycomb lattice the components of $F_1$ order are specified by two vectors $\vet{w}_A$ and $\vet{w}_B$ which collect the order-parameter components for each of the two sublattices (see Ref.~\onlinecite{venderbos15} for details). To construct the scalar triplet state, we pair each component with a distinct spin component, leading to 
\begin{gather} \label{eq:honchiral}
\order{\crea{i\sigma}{\vet{k}+\vet{M}_\mu }\anni{j\sigma' }{\vet{k}}} =  \Delta_{A_2} \;  w^\mu_i\delta_{ij} \sigma^\mu_{\sigma\sigma'}. 
\end{gather}
The vectors $\vet{w}_A$ and $\vet{w}_B$ are given by $\vet{w}_A=(-1,-1,1)$ and $\vet{w}_B=(1,-1,-1)$. 

These two triangular and honeycomb triple-$M$ spin density wave states are examples of non-coplanar spin order. In fact, these spin density waves, which we have derived from symmetry principles here, are nothing but the chiral spin density waves found both in itinerant classical magnets and mean-field Hubbard model calculations on the triangular~\cite{martin08,akagi10,kumar10,venderbos12}, honeycomb~\cite{li12,nandkishore12b,jiang15} and kagome lattices \cite{chern12,barros14,ghosh14}. The chiral spin density wave was also found in a honeycomb Hubbard model study using advanced quantum many-body techniques~\cite{wang12,kiesel12,jiang14}. 

Chiral spin density waves owe their name to the property of having nonzero chirality $\kappa$, defined as $\kappa = \vet{\Delta}_1 \cdot \vet{\Delta}_2 \times \vet{\Delta}_3 \neq 0$ [where $\vet{\Delta}_\mu \sim \sum_k \order{\crea{}{\vet{k}+\vet{M}_\mu }\vet{\sigma}\anni{  }{\vet{k}}}/N ]$. The chiral spin density waves of~\eqref{eq:trichiral} and~\eqref{eq:honchiral} were shown to induce a full spectral gap and the mean-field ground state is a Chern insulator with spontaneous quantum Hall (QH) effect~\cite{martin08,li12}. This is consistent with $A_2$ symmetry, i.e., the breaking of all reflections, and broken time-reversal symmetry caused by non-coplanar spin order. 

\begin{figure}
\includegraphics[width=\columnwidth]{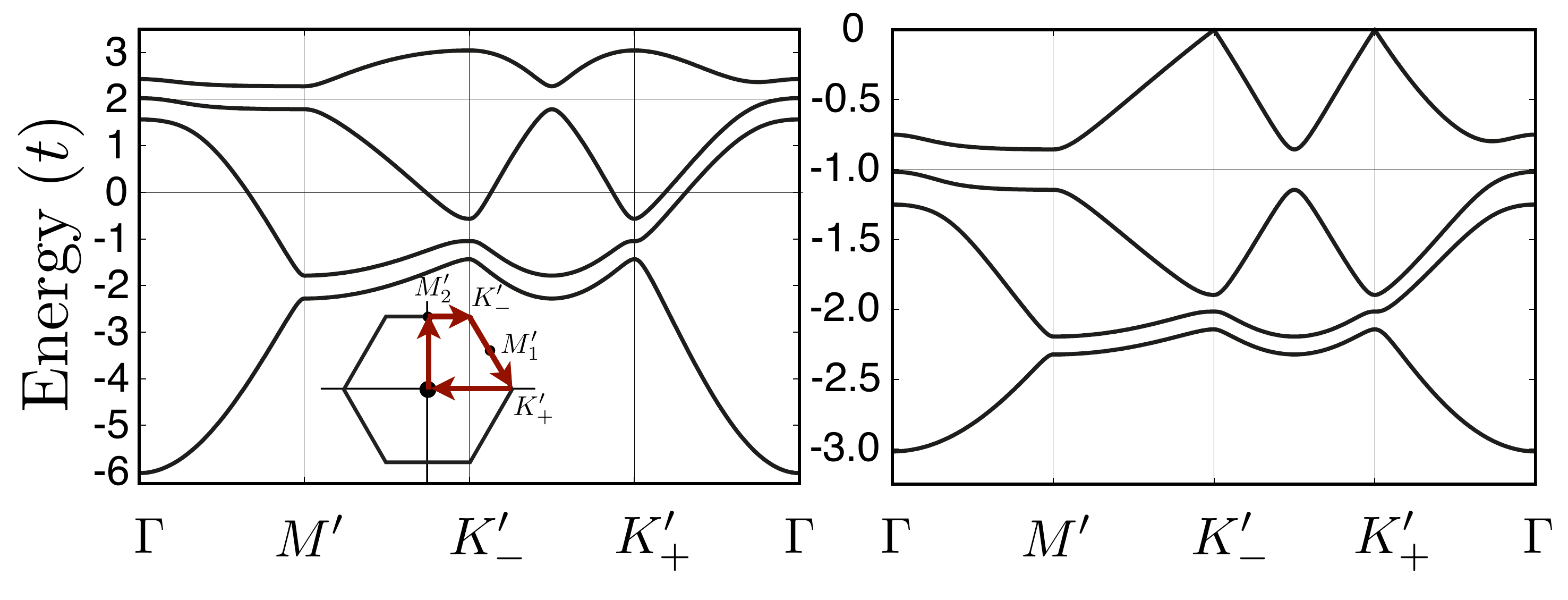}
\caption{\label{fig:swavespec}. Mean-field spectra of the scalar triplet $s$-wave states of the triangular lattice (left) given in Eq.~\eqref{eq:trichiral}, and of the honeycomb lattice (right) given in Eq.~\eqref{eq:honchiral}. In both cases we show spectra for $\Delta_{A_2} = 0.25$. The inset on left shows the reduced BZ with the path along which bands are plotted. Note that all bands are doubly degenerate, as explained in the text. In case of the honeycomb lattice (right) we only show the lower half of the spectrum, i.e., up to $E=0$. At fillings $n=3/4$ (triangular) and $n=3/8$ (honeycomb) the mean field ground state is insulating and has nonzero Chern number.}
\end{figure}

The mean-field spectra of these scalar $s$-wave states are presented in Fig.~\ref{fig:swavespec}. They show the spectral gap at van Hove filling. Note that the spectra exhibit a manifest two-fold degeneracy, i.e., each band is fully two-fold degenerate. This can be traced back to the effective translation invariance of the chiral spin density wave: translations are preserved when followed by a global rotation. As a consequence of $A_2$ symmetry there can be no further degeneracies. To obtain the mean-field spectra we used a tight-binding mean-field Hamiltonian $H_0 + H_\Delta$, where $H_0$ contains a nearest-neighbor hopping $t=1$ and $H_\Delta$ contains the mean-fields defined above.

The scalar triplet $s$-wave states are the spin density waves obtained from mapping the $T_{2g}$ $\beta$-phase back to a density wave. A natural question then is: What is the density wave state corresponding to the $T_{2g}$ $\alpha$-phase of Eq.~\eqref{eq:cubicalpha}? Clearly, this must be a uniaxial spin density wave, so each ($M$-point) component is associated with the same spin matrix (see Fig.~\ref{fig:f1spin}). Let us consider a particular ``$\alpha$ phase'', where each component has equal amplitude $ \vet{\Delta}_{\alpha} \sim \Delta_\alpha (1,1,1)$. In the spin density wave language this is a triple-$M$ uniaxial spin density wave given by
\begin{gather} \label{eq:triuniaxial}
\order{\crea{\sigma}{\vet{k}+\vet{M}_\mu }\anni{\sigma' }{\vet{k}}} =  \Delta_{\text{uniaxial}} \;  \sigma^3_{\sigma\sigma'},
\end{gather}
with uniform order parameter magnitude $\Delta_{\text{uniaxial}}$ but $\kappa=0$. This state is the uniaxial spin density wave of Ref.~\onlinecite{nandkishore12b}. 

The connection between the uniaxial triplet states and the chiral triplet states has been demonstrated from the perspective of their mean-field spectra~\cite{chern12}. The former are semi-metallic with a (spin-filtered) quadratic band crossing at the reduced zone center~\cite{nandkishore12b}. Smoothly deforming the uniaxial spin state so as to give it nonzero spin chirality $\kappa$ gaps out this quadratic band crossing point and results in a state adiabatically connected to the gapped scalar spin-orbit coupled state of Eq.~\eqref{eq:trichiral}~\cite{chern12}. 

We thus find that by mapping the cubic spin-orbit coupled liquid crystal phases back to density wave states, we are naturally led to the chiral and uniaxial spin density wave states. The upshot of this mapping is that their symmetries are made transparent.

\subsection{$d$-wave triplet states: Spin-current density waves\label{ssec:dwave}}

The triplet $d$-wave states have $d$-wave form factors associated with each of the three ordering components $\vet{M}_\mu$. As a consequence, they are imaginary, preserve time-reversal symmetry, and are described by the nesting instability matrices $\vet{\Lambda}_b\sigma^j$ in Eq.~\eqref{eq:nestsym}. We found that the scalar triplet order constructed from $d$-wave components has $A_1$ symmetry: it is the $T_{1g}$ $\beta$ phase mapped back to a density wave state. 

It is obtained by pairing each $d$-wave component at wave vector $\vet{M}_\mu$ with a different spin component $\sigma^j$. In case of the triangular lattice, writing the condensate in terms of $\Delta_\mu(\vet{k})$ as $\order{\crea{\sigma}{\vet{k}+\vet{M}_\mu }\anni{\sigma' }{\vet{k}}} =  [\Delta_\mu(\vet{k})]_{\sigma\sigma'}$, we find
\begin{gather}
\Delta_1(\vet{k})  =  i \Delta_{A_{1}}  (\cos k_3 - \cos k_1 )\sigma^1, \nonumber \\
\Delta_2(\vet{k})  =  i \Delta_{A_{1}}  (\cos k_1 - \cos k_2 ) \sigma^2, \nonumber \\
\Delta_3(\vet{k})  =  i \Delta_{A_{1}}  (\cos k_2 - \cos k_3 ) \sigma^3 . \label{eq:trispinf2}
\end{gather}
Here, $\Delta_{A_{1}} $ is a real order parameter and $k_i =
\vet{k}\cdot \vet{a}_i$ with $\vet{a}_i$ three (triangular) lattice vectors related by three-fold rotations. These triplet density waves preserve time-reversal symmetry, since the combination of complex conjugation and spin flip leaves the state invariant. In that sense, it represents genuine dynamically generated spin-orbit coupling and is therefore the time-reversal invariant analog of the chiral spin density wave. 

We note that this is different for the cubic equivalent of the scalar hexagonal $d$-wave state, i.e., the $\beta$ phase of $L=4$ $T_{1g}$ orbitals. Collecting the $T_{1g}$ orbitals listed in Table~\ref{tab:correspond} in the vector $\vet{g}(\vet{k})$, the cubic $L=4$ $\beta$ phase takes the form
\begin{gather} \label{eq:cubicbeta2}
\order{\hat{\chi}^\dagger_{\sigma}(\vet{k})\hat{\chi}_{\sigma' }(\vet{k})} =  \Delta_{\beta} \; \vet{g}(\vet{k}) \cdot \vet{\sigma}_{\sigma\sigma'}.
\end{gather}
Hermiticity requires $  \Delta_{\beta}$ to be real, implying that the cubic spin-orbit coupled $\beta$ phase and the scalar $M$-point $d$-wave state are equivalent with respect to all spatial symmetries, yet differ with respect to time reversal. 

\begin{figure}
\includegraphics[width=\columnwidth]{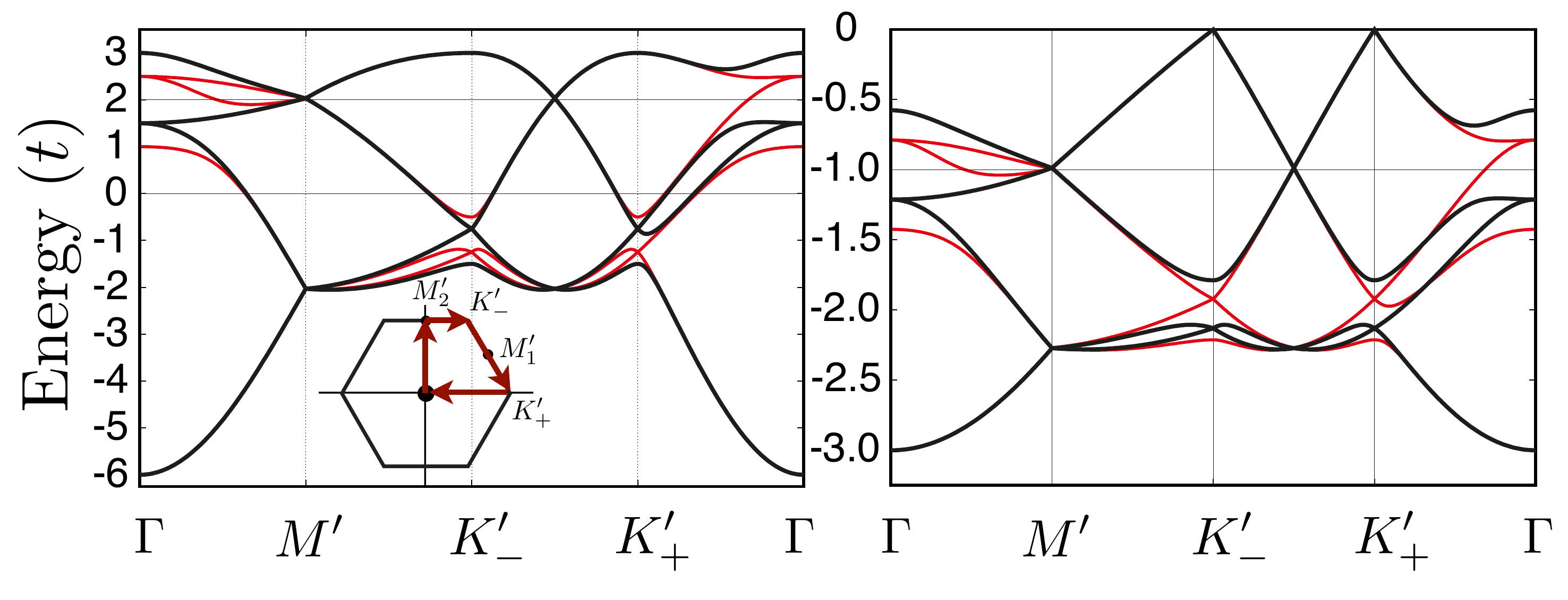}
\caption{\label{fig:dwavespec}. Mean-field band structure of the scalar triplet $d$-wave state of Eq.~\eqref{eq:trispinf2} (left), and the equivalent state on the honeycomb lattice (right). Inset on the left shows the path taken in the reduced BZ. On the left we plot $\Delta_{A_{1}}=0.25$ (black) and $\Delta_{A_{1}}=-0.25$ (red), whereas on the right we plot $\Delta_{A_{1}}=\pm 0.15$ (black/red). All bands are two-fold degenerate. The key feature of the $d$-wave band structures shown here is the symmetry protected degeneracy at the $M'$ points, and the resulting the Dirac semimetal mean-field state. }
\end{figure}
 
Let us study the mean-field spectrum of the density wave state of Eq.~\eqref{eq:trispinf2}, which is shown in Fig.~\ref{fig:dwavespec}. We first note a full two-fold degeneracy of each band, as was the case for the scalar $s$-wave state. The reason for the degeneracy is the same: translations (in combination with spin rotations) are good symmetries. In the present case, however, the presence of both time-reversal symmetry and inversion also mandates a two-fold degeneracy, which will therefore be preserved even if translations are broken. We further observe that the spectrum depends on the sign of $\Delta_{A_{1}}$, as the black (red) curves correspond to positive (negative) sign.  
For $\Delta_{A_{1}}>0$ (black bands), we find that the mean-field ground state is a semimetal, with linearly dispersing nodal points located at the $M'$ points of the reduced BZ. Due to the two-fold degeneracy of each band, the Dirac nodes come in pairs \emph{at each} $M'$ point, giving rise to three flavors of four-component Dirac nodes.

These nodal points are protected by crystal symmetries, and as a result the semi-metallic mean-field state is a symmetry-protected Dirac semimetal in two dimensions~\cite{young12,young15}. We show this explicitly in the next section. As such, the scalar triplet $d$-wave state with $A_1$ symmetry should be contrasted with graphene~\cite{neto09}. Graphene has fully spin-degenerate Dirac nodes which can be gapped by symmetry-preserving spin-orbit coupling~\cite{kane05a}. In contrast, in the present case the mean-field Dirac quasiparticles can only become massive by breaking symmetries, leading to either a trivial insulating state or a topological insulator. Therefore, the Dirac semimetal state originating from scalar triplet $d$-wave state sits at the boundary between a trivial and topological insulator~\cite{young15}. 

The symmetry protected 2D Dirac semimetal is a generic feature of triplet $M$-point order, since it is rooted in hexagonal symmetry. This is confirmed by Fig.~\ref{fig:dwavespec}, which shows the mean-field spectrum of the honeycomb lattice scalar triplet $d$-wave state. The key characteristics of the honeycomb lattice spectrum are identical to the triangular lattice spectrum, notably the Dirac nodes at the $M'$ points.

\begin{figure}
\includegraphics[width=\columnwidth]{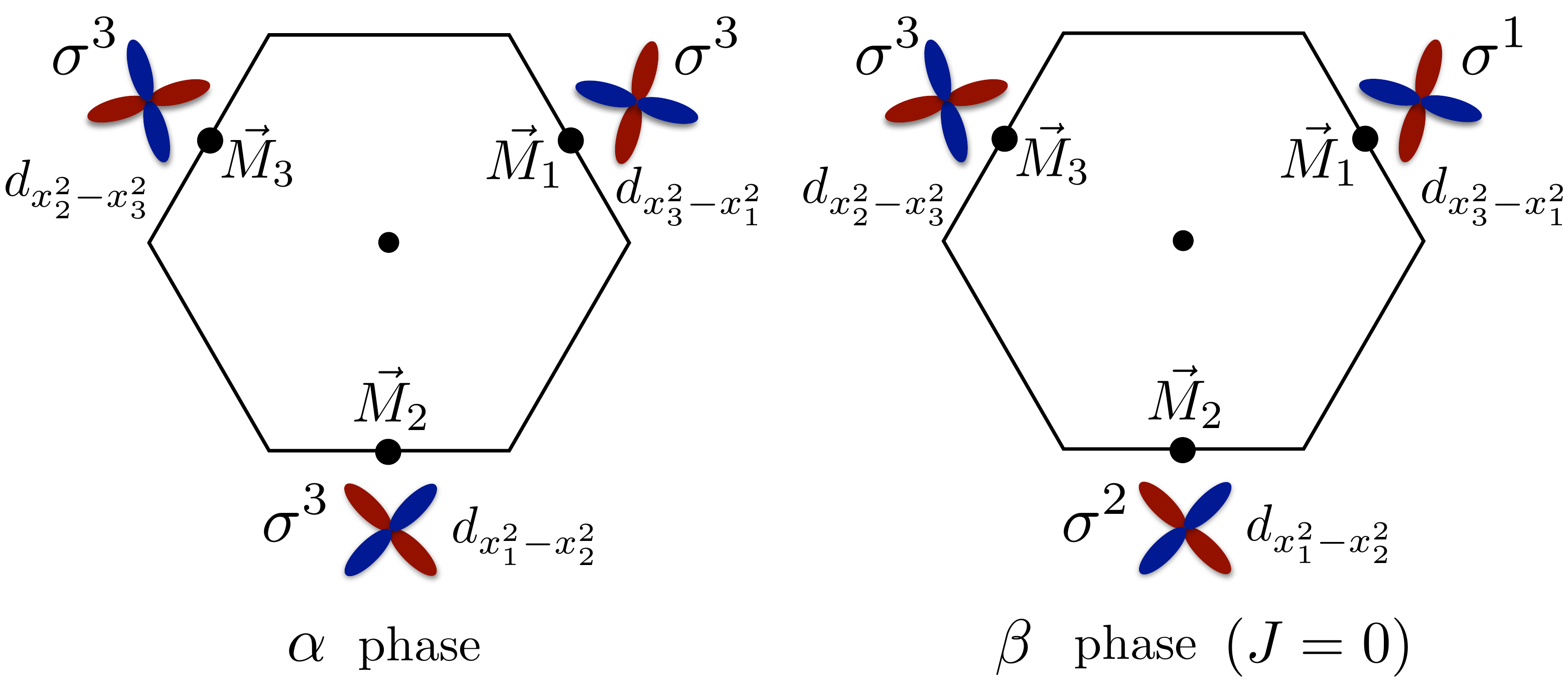}
\caption{\label{fig:f2spin} Schematic representation of $F_2$ $d$-wave triple-$M$ order. The $M$-points are associated with $d$-wave orbitals rotated by $\pm 2\pi/3$ with respect to each other. (Left) Each $M$-point is associated with the same spin matrix giving a uniaxial QSH state. (Right) Each $M$-point is associated with a different spin matrix, leading to the scalar triplet $d$-wave state with Dirac semimetallic spectrum. The pairings on the left and right may be interpreted as cubic $\alpha$ and $\beta$ phases, respectively. }
\end{figure}

Hexagonal lattice triplet $d$-wave order is schematically shown and summarized in Fig.~\ref{fig:f2spin}. On the right side, each wave vector $\vet{M}_\mu$ is paired with its corresponding $d$-wave form factor and a different spin matrix. This depicts spin-orbit coupling. On the left each wave vector is paired with its $d$-wave form factor, however, each wave vector carries the same spin. We refer to this as uniaxial order, in analogy with $s$-wave order in Fig.~\ref{fig:f1spin}. The uniaxial order corresponds to an $L = 4$ cubic $\alpha$ phase, similar to the $L = 2$ $\alpha$ phase of Eq.~\eqref{eq:cubicalpha}. The $\alpha$-phase is interesting in its own right, since it corresponds to a quantum spin Hall (QSH) phase~\cite{kane05a,kane05b}. It can be viewed as two copies of a $d$-wave Chern insulator, one for each spin species with opposite sign. 

\section{Low-energy properties of hexagonal triplet orders\label{sec:lowenergy}}

The aim of this section is to give a more elaborate analysis of the spectral properties of states discussed in the previous section. In particular, we examine to what extent the characteristics of the mean-field ground state can be understood from a low-energy description. Such a description is independent of a given lattice model and therefore helps to put the result on a more general footing. First, we study the lifting of energy level degeneracies at the $M$ points where the van Hove electrons are located. We then derive the low-energy Dirac theory of the nodal degeneracies that arise in the triplet $d$-wave state of Eq.~\eqref{eq:trispinf2}. 

\subsection{Electrons at the $M$ points\label{ssec:mpockets}}

We start from the electron operator $\hat{\Phi}$ of Eq.~\eqref{eq:vanhovehexa} and consider the action of symmetries on $\hat{\Phi}$. The action of the symmetry group $C'''_{6v}$ on the three flavors of $M$-point electrons is given by the $M$-point representation matrices $\{ G_i,X,Y \} $ introduced in the previous section (see also Appendix~\ref{app:malgebra}). Specifically, for the symmetry group generators we have
\begin{align} 
T(\vet{a}_1) \; : \;  \hat{\Phi} &\rightarrow \; G_1 \hat{\Phi}, \nonumber \\
C_6 \; : \;  \hat{\Phi} &\rightarrow \; X \hat{\Phi} \nonumber \\
\sigma_v \; : \;  \hat{\Phi} &\rightarrow \; Y \hat{\Phi}.  \label{eq:vanhovesym}
\end{align}
Density wave ordering is expressed in terms of fermion bilinears, $\hat{\Phi}^\dagger \Lambda^i\sigma^j \hat{\Phi}$, as discussed in the beginning of the previous section. 

In case of singlet (spin-rotation invariant) order, i.e. $\hat{\Phi}^\dagger \Lambda^i\hat{\Phi}$, the transformation properties of the $M$-point electrons can be used to show that the set of Gell-Mann matrices $\vet{\Lambda}_a$ has $F_1$ symmetry and the set $\vet{\Lambda}_b$ has $F_2$ symmetry~\cite{venderbos15}. Based on that, we found that the symmetry of triple-$M$ order, ordering of each of the $M$-point components simultaneously with equal amplitude, is $A_1$ for the former and $A_2$ for the latter. The energy levels of triple-$M$ order are given by the two Gell-Mann matrices $\Lambda^2_c$ and $\Lambda^1_c$, respectively, in the corresponding eigenbasis.

The effect of spin structure is best accounted for by explicitly distinguishing the two types of triplet order, uniaxial and spin-orbit coupled, and analyzing them separately. In the first case, that of uniaxial triplet order, the analysis is a straightforward extension of singlet order~\cite{venderbos15}. In the second case, that of scalar spin-orbit coupled order with full SU(2) symmetry breaking, the analysis of energy level splittings at the $M$ points is more subtle and requires the notion of double groups. It turns out that, as a consequence of the intimate connection to cubic symmetry, energy levels are governed by the double cubic group, as we will explain in what follows.

Let us first focus on the uniaxial triplet orders. To be specific, we take the spin polarization axis to be the $z$ axis. $M$-point electron bilinears can then be written as a simple product of Gell-Mann matrices and $\sigma^3$. In particular, the two bilinears describing uniaxial triple-$M$ order are given by $\Lambda^2_c\sigma^3$ (uniaxial spin density wave) and $\Lambda^1_c\sigma^3$ (uniaxial orbital spin currents)~\cite{note1}. As a result, the analysis of the spin-rotation invariant case effectively applies to each spin sector separately. For the uniaxial spin density waves of Eq.~\eqref{eq:triuniaxial}, this implies a two-fold degeneracy in each spin sector. However, due to the relative sign difference ($\sim \sigma^3$), these two-fold degenerate levels in each spin sector are split with respect to each other. In addition, there are two non-degenerate levels, one for each spin species. Importantly, both spin-filtered two-fold degeneracies constitute a spin-filtered quadratic band touching (QBT) protected by rotational symmetry and an effective time-reversal symmetry~\cite{sun09,nandkishore12b,chern12}. This directly follows from the symmetry of triple-$M$ order~\cite{venderbos15,note2}. Furthermore, since such argument only relies on symmetry, it proves that the ``half-metal'' state of Ref.~\onlinecite{nandkishore12b}, i.e., a metal with fully spin-polarized Fermi surface electrons, is a generic feature of uniaxial $M$-point spin density waves.

\begin{table*}[t]
\centering
\begin{ruledtabular}
\begin{tabular}{cccccccc}
Rep. & Type  &  triple-$M$ &  GS  & triple-$M$ & GS & triple-$M$& GS \\ 
&  & singlet &  &  uniaxial &  & SOC scalar&  \\ 
\hline
$F_1$  & $s$-wave & $A_1\; (+,-)$ & insulator/QBT  & $A_1\;(-,-)$ & spin-filtered QBT & $A_2\;(-,+)$ & Chern insulator \\
$F_2$  & $d$-wave & $A_2\; (-,-)$ & Chern insulator  & $A_1\;(+,-)$ & QSH insulator & $A_1\;(+,+)$ & spin-locked Dirac SM \\
\end{tabular}
\end{ruledtabular}
 \caption{Summary of hexagonal lattice $s$- and $d$-density wave states at the $M$-points. Table lists the symmetry and nature of the mean-field ground state (GS) of singlet triple-$M$, uniaxial triple-$M$, and spin-orbit coupled (SOC) scalar triple-$M$ order. The symmetry label $A_{1,2}$ refers to point group symmetry, and $(\pm,\pm)$ refers to preserved/broken time-reversal (first entry) and translational symmetry (second entry). Note that in case of uniaxial order we give the label $A_1$ since one can consider each spin species separately, and invert the spin if necessary. }
\label{tab:msummary}
\end{table*}

The gap matrix $\Lambda^1_c\sigma^3$, which corresponds to orbital spin currents, leads to a double degeneracy of all three energy levels of $\Lambda^1_c$: each level occurs once for each spin projection. The matrix $\Lambda^1_c$ implies time-reversal symmetry breaking, but in combination with $\sigma^3$ time-reversal symmetry is preserved. Moreover, since the gap matrix $\Lambda^1_c$ corresponds to a spontaneous QH phase, $\sigma^3$ promotes the ground to a QSH phase~\cite{kane05a,kane05b}.

Next, we consider the case of spin-orbit coupled scalar order. Contrary to uniaxial order, spin-orbit coupled order does not decouple into two separate spin sectors. For instance, scalar $s$-wave order of Eqs.~\eqref{eq:trichiral} and \eqref{eq:honchiral} is written in terms of van Hove electron bilinears as $\vet{\Lambda}_a \cdot \vet{\sigma}$. As a result, one needs to consider the combined effect of symmetries on spin and $M$-point degrees of freedom. We now show that degeneracies in the subspace given by $\hat{\Phi}$ can be derived by exploiting the mapping between hexagonal triplet $M$-point order and cubic orbital order. 

To demonstrate that the energy levels of van Hove electrons $\hat{\Phi}$ are effectively governed by the cubic double group, we construct an operator $\hat{\Upsilon} $ so that its orbital degree of freedom transforms in the same way under $O_h$ as $\hat{\Phi}$ under $C'''_{6v}$. The operator $\hat{\Upsilon} $ is given by the $T_{2g}$ orbitals, $\hat{\Upsilon}  = (\hat{\psi}_{yz\sigma} , \hat{\psi}_{zx\sigma},\hat{\psi}_{xy\sigma} )$ (see also Appendix~\ref{app:malgebra}). For instance, under a two-fold rotation about the $z$-axis (equivalent to the translation $T(\vet{a}_1)$ in $C'''_{6v}$) $\hat{\Upsilon} $ transforms as $G_1 \hat{\Upsilon} $. Similarly, other elements of $O_h$ act on $\hat{\Upsilon} $ as products of $\{ G_i,X,Y \} $, the generators of the $M$-point representation.  

With spin-orbit coupling symmetries act on $\hat{\Upsilon} $ as double group elements. The action of symmetries such as the rotation $G_1$ is then $U_{G_1} G_1 \hat{\Upsilon} $, where the $SU(2)$ matrix $U_{G_1} $ implements the rotation $G_1$ in spin space. In general, the matrix $U_g$ implements the symmetry $g \in O_h$. Symmetry-mandated degeneracies of $\hat{\Upsilon} $ follow from representations of the double group. The cubic double group admits 2D and 4D spin-orbit coupled representations, corresponding to total angular momenta $j=1/2$ and $j=3/2$. It is known that under cubic symmetry the $T_{2g}$ orbitals split into a two-fold degenerate $j=1/2$ doublet and a four-fold degenerate $j=3/2$ quadruplet. This situation applies to the scalar order with $A_{1g}$ symmetry (i.e., symmetric under all elements of the cubic group) of Eq.~\eqref{eq:cubicorder}. Instead, the scalar order with $A_{2g}$ symmetry is symmetric under all elements of $T_h$. With spin-orbit coupling the double group of $T_h$ only admits $2D$ representations, and as a result degeneracies will be two-fold. 

The key result is that degeneracies of $\hat{\Upsilon}$ carry over to $\hat{\Phi}$. The splitting of $M$-point electrons is identical to the splitting of the $T_{2g}$ orbitals. This is a direct consequence of the mapping between cubic and hexagonal $C'''_{6v}$ symmetry: symmetries acting on $\hat{\Upsilon}$ must act in the same way on $\hat{\Phi}$ and therefore degeneracies are preserved.
The mean-field spectra of scalar triplet $M$-point order confirm this. The spectra of the scalar triplet $d$-wave states shown in Fig.~\ref{fig:f2spin} exhibit a two-fold and four-fold degeneracy at $\Gamma$, the order of which depends on the sign of $\Delta_{A_1}$ (i.e., black and red bands). Instead, all levels at $\Gamma$ are two-fold degenerate in case of scalar $s$-wave triplet order shown Fig.~\ref{fig:f1spin} (recall that all bands are two-fold degenerate).  

In addition to explaining degeneracies, the electron operator $\hat{\Upsilon}$ gives rise to a dual description of the mapping from hexagonal density waves to nonzero angular momentum condensation in $3D$. Instead of associating the angular momentum with the condensed particle-hole pairs, as we have done so far, it can be associated with an internal electronic orbital degree of freedom given by the $T_{2g}$ orbitals of $\hat{\Upsilon}$. Condensation in the $s$-wave ($L=0$) channel then corresponds to spontaneous orbital order, which is expressed by the bilinears 
\begin{gather}
\order{\hat{\Upsilon}^\dagger \vet{\Lambda}_a\sigma^j \hat{\Upsilon}} , \quad \order{\hat{\Upsilon}^\dagger \vet{\Lambda}_b\sigma^j\hat{\Upsilon}}.
\end{gather}
Since $\hat{\Upsilon}$ and $\hat{\Phi}$ transform equivalently under cubic and hexagonal symmetry, respectively, these orbital order parameters are symmetry-equivalent to the spin density and spin-current density waves. 


The considerations based on a description in terms of low-energy $M$-point electrons are summarized in Table~\ref{tab:msummary}. The two sets of $M$-point order components, $s$ wave ($F_1$ symmetry) and $d$ wave ($F_2$ symmetry), can condense in singlet or triplet channel. In the latter case, assuming triple-$M$ ordering, there is a uniaxial phase and a spin-orbit coupled scalar phase. In the uniaxial state translational symmetry is broken and the mean-field ground state is a spin-filtered QBT semimetal or a QSH insulator. In the scalar spin-orbit coupled state translational symmetry is preserved and the mean-field ground state is a Chern insulator or a symmetry protected Dirac semi-metal.

\subsection{Low-energy theory at the $M'$ point: 2D Dirac semimetal\label{ssec:dirac}}

The main characteristic of the scalar triplet $d$-wave state is the nodal Dirac degeneracy at the $M'$ points of the reduced BZ, as shown in Fig.~\ref{fig:dwavespec}. We argued that these Dirac points are symmetry protected and we will now prove this. To this end we choose the $\vet{M}'_2$ point and write the electron operator at $\vet{M}'_2$ (in case of the triangular lattice) as
\begin{gather} \label{eq:m2operator}
\hat{\Phi}_{M'}  \equiv  \begin{pmatrix} \anni{\sigma}{\vet{M}'_2} \\  \anni{\sigma}{\vet{M}'_2+\vet{M}_1} \\   \anni{\sigma}{\vet{M}'_2+\vet{M}_2} \\  \anni{\sigma}{\vet{M}'_2+\vet{M}_3}    \end{pmatrix}.
\end{gather}
To facilitate expressing the action of lattice symmetries, we define a set of Pauli matrices $\tau^i$ acting \emph{within} the blocks $(\vet{M}'_2,\vet{M}'_2+\vet{M}_1 )$ and $(\vet{M}'_2 +\vet{M}_2,\vet{M}'_2+\vet{M}_3 )$, in addition to a set of matrices $\nu^i$ acting on the block degree of freedom (e.g., exchanging blocks). The $\vet{M}'_2$ point is left invariant by the inversion $C_2$, the reflection $\sigma_v$, and notably the translations $T(\vet{a}_i)$, all of which are symmetries of the scalar triplet $d$-wave state (in combination with global spin rotations). 

We find that the inversion acts as $\nu^1 \hat{\Phi}_{M'} $, whereas the translations $T(\vet{a}_2)$ and $T(\vet{a}_3)$ act as $\sigma^1\nu^3 \hat{\Phi}_{M'} $ and $\sigma^2\tau^3 \hat{\Phi}_{M'} $ (see Appendix~\ref{app:dirac}). The Hamiltonian matrix at $\vet{M}'_2$ is linear combination of matrices $\sigma^i\nu^j\tau^k$ ($i,j,k=0,1,2,3$) and must be invariant under the symmetry operations. Taking into account the constraints coming from $C_2\sigma_v$ and time-reversal $\mathcal{T}$ we find only two allowed terms at $\vet{M}'_2$, which are $\tau^3 $ and $\tau^2(\sigma^1 - \sigma^3\nu^1)$. These two terms anti-commute, implying two eigenvalues at $\vet{M}'_2$, $\epsilon$, and $-\epsilon$, each four-fold degenerate. We conclude that the double Dirac node at $\vet{M}'_2$, and thus at all $M'$ points, is mandated by symmetry. 

\begin{figure}
\includegraphics[width=0.9\columnwidth]{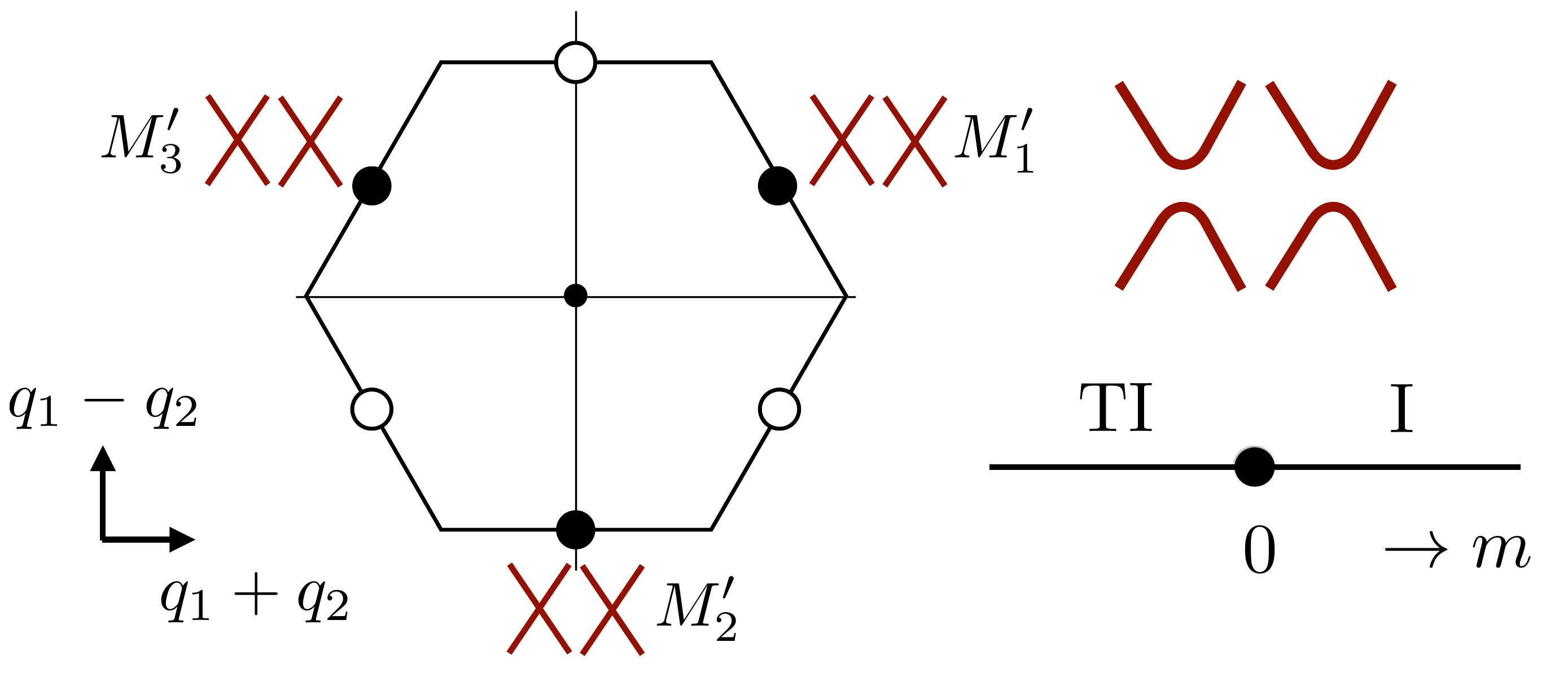}
\caption{\label{fig:diractheory} (Left) Schematic representation of the double Dirac nodes of the scalar triplet $d$-wave state of Eq.~\eqref{eq:trispinf2} and Fig.~\ref{fig:dwavespec}, located at the three \emph{inequivalent} $M'$ points. (Right) The Dirac points can be gapped out by a charge density modulation [Eq.~\eqref{eq:perturb}] controlled by mass parameter $m$, the sign of which determines whether gapped state is a topological insulator (TI) or a trivial insulator (I). }
\end{figure}

Based on this conclusion, we ask whether symmetry breaking perturbations can gap out the Dirac nodes in interesting ways. First, we expand the mean-field band structure corresponding to the Hamiltonian $H_0 + H[\Delta]$ [with $\Delta \equiv \Delta_{A_1}$ of Eq.~\eqref{eq:trispinf2}] around $\vet{M}'_2$, assuming we are in the semi-metallic state shown in Fig.~\ref{fig:dwavespec} (black bands). We take this node as an example, the theory around $\vet{M}'_{1,3}$ may be developed in a similar way. Details of the derivation are presented in Appendix~\ref{app:dirac}, and we simply quote the result here:
\begin{gather} \label{eq:doubledirac}
\mathcal{H}(\vet{q}) = (v_1 q_- +v'q_+)\tilde{\nu}^1\tilde{\tau}^2+(v_2 q_+ -v'q_-)\tilde{\nu}^3\tilde{\tau}^2.
\end{gather}
The Pauli matrices $\tilde{\nu}^i$ and $\tilde{\tau}^i$ act on an effective valley and pseudospin degree of freedom of the double Dirac node, specified in Appendix~\ref{app:dirac} together with Fermi velocity coefficients $v_{1,2},v'$. We have defined $q_\pm = q_1 \pm q_2$, where $q_i = \vet{q}\cdot \vet{a}_i$. As shown in Fig.~\ref{fig:diractheory}, $q_+ \sim q_x$ and $q_- \sim q_y$, implying that $q_-$ is along the direction of the undistorted Fermi surface (see Fig.~\ref{fig:bzhexa}), whereas $q_+$ is orthogonal to it. We expect that when $\Delta_{A_1}\rightarrow 0 $ the Hamiltonian $\mathcal{H}(\vet{q}) $ only depends on $q_+$. This is verified by checking the behavior of $v_{1,2}$ and $v'$, which become $v_1,v' \rightarrow 0$, and $v_2 \rightarrow 2t$. As a result, the Hamiltonian~\eqref{eq:doubledirac} describes the double Dirac node with linear dispersion in $q_\pm$ as function of the order parameter $\Delta_{A_1}$.

The symmetry protection of the double Dirac node (at $\vet{M}'_2$) critically relies on the invariant translations $T(\vet{a}_i)$. To study the fate of the Dirac node, we therefore consider a perturbation that breaks translational symmetry. Such a perturbation is given by the charge modulation
\begin{gather} \label{eq:perturb}
\delta\mathcal{H} = m \sum_{\mu,\sigma,\vet{k}} \crea{\sigma}{\vet{k} }\anni{\sigma}{\vet{k}+\vet{M}_\mu} + \text{H.c.},
\end{gather}
and we find that $\delta\mathcal{H}$ gaps out the Dirac nodes at the $M'$ points. The gapped state respects time-reversal and inversion symmetries, and we calculate the Fu-Kane invariant $\nu_0$~\cite{fu07} to determine the nature of the ground state. Quite remarkably, we find that the topological invariant $\nu_0$ depends on the sign of $m$, i.e., $(-1)^{\nu_0} = \text{sgn}(m)$. This result may be understood as follows, starting from double Dirac node at $\vet{M}'_2$. The double Dirac node consists of two Kramer's doublets with opposite inversion eigenvalues. The time-reversal invariant perturbation $\delta\mathcal{H} $ splits the double node, leaving only one of the Kramer's doublets occupied. The sign of $m$ controls which Kramer's doublet, and consequently which inversion eigenvalue, is occupied. If the \emph{even} eigenvalue is occupied at $\vet{M}'_2$, the same must be true for the other $M'$ points, implying that the product of inversion eigenvalues of occupied bands at time-reversal invariant momenta, and thus $\nu_0$, is \emph{odd} (since the product of the $-\epsilon$ subspace is odd). This shows that $\text{sgn}(m)$ determines whether the gapped state is a topological or trivial insulator. In Appendix~\ref{app:dirac}, we offer an alternative interpretation of this result. 

The main features of the mean-field Dirac semi-metal state are summarized in Fig.~\ref{fig:diractheory}. In particular, Fig.~\ref{fig:diractheory} highlights that the Dirac semimetal sits at the boundary between a trivial and topological insulator. Both phases are accessible by a single perturbation parameter $m$, given by Eq.~\eqref{eq:perturb}. We stress that this applies in general to lattices with hexagonal symmetry. We leave a comprehensive investigation of the double Dirac node theory, including a classification of all possible mass terms, for future study. 

\section{General symmetric spin states\label{sec:classical}}

The symmetry classification of spin density waves has a connection to a special class of classical spin states, the symmetric classical spin states. Symmetric classical spin states are configurations of classical spins that respect all symmetries of the crystal lattice, up to a global $O(3)$ rotation. The concept of symmetric classical spin states was introduced in Ref.~\onlinecite{messio11}, where they were referred to as \emph{regular magnetic orders}. The significance of the global spin rotations lies in the fact that most spin Hamiltonians are functions of $O(3)$ invariant bilinears such as  $\vet{S}_i \cdot \vet{S}_j$. In Ref.~\onlinecite{messio11} it was shown that symmetric classical spin states are good variational classical ground states of those spin Hamiltonians.

The scalar triplet $s$-wave states of~\eqref{eq:trichiral} and~\eqref{eq:honchiral} are examples of such symmetric spin states, if we interpret them as classical spin states. Formally, we can identify these triplet density waves with classical spin configurations $\vet{S}(\vet{x}) = \sum_{\vet{q}}  \vet{S}(\vet{q}) e^{-i\vet{q}\cdot \vet{x}}$ by taking the Fourier components $\vet{S}(\vet{M}_\mu) $ equal to 
\begin{gather} \label{eq:classical}
S^i(\vet{M}_\mu)  = \frac{1}{N}\sum_{\sigma\sigma'\vet{k}} \sigma^i_{\sigma\sigma'}\order{\crea{\sigma}{\vet{k}+\vet{M}_\mu }\anni{\sigma' }{\vet{k}}}.
\end{gather}
(For simplicity we have suppressed the sublattice index.) The resulting spin configuration respects all lattice symmetries up to global spin rotation. To see this, we recall that the scalar triplet states transform as scalars under $C'''_{6v}$, including the translations, since global spin rotation invariance is a manifest feature of the classification of spin density waves. The scalar triplet states are invariant up to sign, which implies that the classical spin states derived from them are indeed symmetric.

We can ask the following question: Can we obtain all symmetric spin states with $M$-point modulation on a given hexagonal lattice using the symmetry classification? The answer is yes. To demonstrate this, it is helpful to review how the scalar triplet orders of Eqs.~\eqref{eq:trichiral} and~\eqref{eq:honchiral} are constructed. The starting point is three-component charge order with $F_1$ symmetry, which is then identified with $T_{2g}$ orbital angular momenta in a cubic crystal. Coupling the ($\sim L=1$) angular momenta to spin ($\sim S=1$) and decomposing them into total angular momentum states yields an effective $J=0$ state transforming as a scalar. This scalar non-coplanar spin density wave corresponds to a symmetric configuration of classical spins.

Apart from charge order with $F_1$ symmetry, hexagonal lattices can support other charge ordered states with $M$-point modulations (the honeycomb lattice is an example), with different symmetry~\cite{venderbos15} (we briefly review this in Appendix~\ref{app:classical}). Since representations of $C'''_{6v}$ map to representations of the cubic group, distinct charge orders map to distinct angular momenta, such as $p$ or $f$ orbitals, which are three-fold degenerate in cubic symmetry. Coupling these angular momenta to spin, as discussed in Section~\ref{ssec:soc}, and decomposing into total angular momentum states will result in a scalar $J=0$ state. That state, when transformed back to spin density wave must correspond to a symmetric classical spin state, as it is invariant under all lattice symmetries up to a global spin rotation. 

We illustrate this using the honeycomb and kagome lattices as examples. In addition to charge order with $F_1$ symmetry, the honeycomb lattice supports charge order with $F_4$ symmetry. The equivalent of $F_4$ symmetry in the cubic group is $T_{2u}$ symmetry (i.e., $f$ orbitals), and spin-orbit coupling yields $T_{2u}\times T_{1g} = A_{2u} + \ldots $ (we are only interested in the scalar representation). The $A_{2u}$ term implies the existence of a triplet density wave state with $B_1$ symmetry, which corresponds to a symmetric spin state using Eq.~\eqref{eq:classical}. As a result, the honeycomb lattice admits two symmetric classical spin states with $M$-point wave vectors. Applying the same method to the kagome lattice we find three sets of $M$-point charge order: $F_1$, $F_3$, and $F_4$ symmetries (see Appendix~\ref{app:classical}). These are mapped to $d$-, $p$- and $f$-wave angular momenta, respectively. Spin-orbit coupling yields three distinct scalar $J=0$ states, from which we conclude that the kagome lattice admits three symmetric spin states with $M$-point modulations. This is summarized in Table~\ref{tab:classical}. Explicit comparison of the spin configurations shows that our result is in agreement with Ref.~\onlinecite{messio11}.

The constructive derivation presented here provides a straightforward route to obtain the set of symmetric spin states of a given lattice. It is limited only in the sense that the modulation wave vectors are fixed ahead of time. Here we explicitly constructed $M$-point symmetric spin states. Consequently, symmetric spin states with $K$-point modulation, for instance, are inevitably missed. This may be remedied, however, by simply deriving all $K$-point charge order representations and associating them with spin similar in spirit to spin-orbit coupling. Indeed, in case of the kagome lattice, symmetric spin states with $K$-point wave vector can be extracted from $K$-point charge order. 

Symmetric classical spin states are good variational ground states of a large class of spin Hamiltonians, in some cases saturating rigorous lower bounds on ground state energies~\cite{messio11}. Apart from lattice magnets described by spin Hamiltonians, symmetric spin states are also relevant for materials described by itinerant carriers coupled to localized (classical) spins. For instance, magnetic states on the kagome lattice, stabilized by itinerant electron-mediated interactions at specific electron densities~\cite{barros14,ghosh14}, were found to be the symmetric spin states. 
In general, the itinerant carrier density, tuned to commensurate doping, sets the magnetic modulation wave vectors. Therefore, the present approach is particularly suited for deriving variational magnetic states of coupled spin-electron models, as the ordering vectors are predetermined. An added benefit is that the symmetry of the \emph{electronic} Hamiltonian directly follows from the derivation. 

In this section we have shown that the symmetric spin states correspond to the cubic $J=0$ singlets obtained from spin-orbit coupling. We conclude by noting that the full set of multiplets (i.e., all representations, including the multicomponent representations) may be viewed as an exhaustive symmetry classification of all (classical) spin states on a given hexagonal lattice. 

\begin{table}[t]
\centering
\begin{ruledtabular}
\begin{tabular}{cccc}
 & Triangular  &  Honeycomb  & Kagome  \\ 
\hline
Charge order  & $F_1$  & $F_1+F_4$ & $F_1+F_3+F_4$ \\
Cubic symmetry  & $T_{2g}$ & $T_{2g}+T_{2u}$ &  $T_{2g}+T_{1u}+T_{2u}$\\
Orbitals  & $\{d \}$ & $\{d \}+\{f \}$ &  $\{d \}+\{p \}+\{f \}$\\
Scalar ($J=0$)  & $A_2$ & $A_2 + B_1$& $A_2 + B_2+B_1$\\
\end{tabular}
\end{ruledtabular}
 \caption{List of the $M$-point modulated symmetric classical spin states on the triangular, honeycomb and kagome lattices. Top row shows the symmetry of the charge ordered states they are derived from, and the second row lists the cubic representations corresponding to charge order with $F_i$ symmetry. The angular momenta (i.e., $p,d,f$-waves) transforming as those representations of the cubic group are listed in the third row. The bottom row lists the symmetry of the electronic scalar spin density wave state (i.e., the total angular momentum $J=0$ state) that follows from coupling orbitals and spin. }
\label{tab:classical}
\end{table}


\section{Discussion and conclusion\label{sec:summary}}

In this work we introduced a classification of hexagonal triplet density waves on the basis of a mapping from 2D order at finite wave vector to 3D $Q=0$ order with nonzero angular momentum. The mapping follows from the isomorphism between the extended hexagonal point group $C'''_{6v}$ and the cubic point group $O_h$. Let us summarize a number of key consequences of the mapping to cubic symmetry. Two important results directly follow. First, in order to correctly determine the symmetry of hexagonal triplet $M$-point order, it is necessary to consider composites of spatial symmetries and global spin rotations. These composites are naturally obtained by mapping to cubic symmetry. Global spin rotation equivalence, which, for instance, mandates the double degeneracy of electronic bands, is manifestly built into the classification in terms of cubic $L > 0$ orders. Therefore, the correct symmetry of the hexagonal triplet density waves follows naturally from the mapping to cubic symmetry. This is particularly important since the 2D Dirac semimetal is protected by composite symmetries.  Second, in order to understand the splitting of energy levels in the $\hat{\Phi}$ subspace [see Eq.~\eqref{eq:vanhovehexa}], at $\Gamma$, one needs to invoke the double group of $O_h$, as we demonstrated in Sec.~\ref{ssec:mpockets}. The fact that $\hat{\Phi}$ splits into a $j=1/2$ doublet and $j=3/2$ quadruplet, or three $j=1/2$ doublets, is inextricably linked to the equivalence of $\hat{\Phi}$ and $\hat{\Upsilon}$. 

A third consequence deserving a comment, is that the symmetry equivalence of hexagonal triplet orders and cubic liquid crystal phases implies a common phenomenological Ginzburg-Landau description. As a result, from the perspective of effective theories based on the symmetry of the order parameter, the two seemingly different classes of orders have shared properties.  A thorough survey of the features of such a Landau theory is left for future study, but we expect that the connection between 2D and 3D orders gives rise to additional insight. An interesting aspect which is worth mentioning is that multi-component orders, such as the $M$-point spin and spin-current density waves, can in general give rise to distinct types of composite orders. These composites may order at temperatures above the transition temperature of the primary order, leading to, for instance, nematic or charge density wave order. The latter possibility has been addressed for the case of the uniaxial spin density wave in Ref.~\onlinecite{chern12b}. Condensation of composite order parameters has been studied to great extent in the context of the iron-pnictide materials, with a focus on nematicity~\cite{si08,fang09,xu08,qi09,fernandes10,fernandes12} and more recently charge density wave and vector chiral order~\cite{fernandes16}, the latter constituting a triplet $d$-wave order.

Fourth, the classification introduced in this work leads to one of our key results: the identification of a new set of phases, the time-reversal invariant scalar triplet $d$-wave states. In addition, it follows naturally from classification that the scalar triplet $d$ wave is the time-reversal invariant analog of the chiral spin density wave. The scalar triplet $d$ wave has remarkable symmetry properties: it does not break any spatial symmetries, is time-reversal invariant, but does break spin-rotation symmetry. As such, it bears similarity to spin nematic order, and due to its peculiar symmetry may be called a ``hidden order'' state. 

It is important to mention that, even though these two types of spin-orbit coupled scalar orders can be interpreted as 3D electronic liquid crystal phases, there is a significant difference between the two. Whereas electronic liquid crystal phases exhibit Fermi surface distortions or reconstructions but remain metallic, the electronic (mean-field) states corresponding to the scalar triplet orders are insulating ($s$ wave) and semi-metallic ($d$ wave). Indeed, the two systems (i.e., a 2D nested Fermi surface and a 3D Fermi liquid) are different. Therefore, in spite of the correspondence, the nature of both the normal state and the condensate is different for the two cases. There is, however, a similarity in the following sense. We have seen that the scalar $s$-wave state, i.e., the chiral spin density wave, gives rise to a fully gapped spectrum, whereas the spectrum of the scalar $d$-wave state has nodal degeneracies due to higher symmetry. The cubic $T_{2g}$ $\beta$ phase, in contrast to a $p$-wave (or $T_{1u}$) $\beta$ phase, exhibits nodes. The cubic $T_{1g}$ $\beta$ phase has the same nodes, but as a result of full cubic symmetry has an extra set of nodes: higher symmetry mandates an extra set of nodes. 

The scalar spin density wave and scalar spin-current density waves, i.e., the $\beta$-phase density waves, both constitute topological phases, in the sense that their mean-field band structures are topological. The chiral spin density wave has nonzero Chern number in the ground state. It is an example of so-called topological Mott (Chern) insulators~\cite{raghu08}. The scalar triplet $d$-wave state is a novel type of semimetal. It realizes a dynamically generated Dirac semi-metal in two dimensions, protected by crystal symmetry. Perturbations can gap out the Dirac nodes and drive the system into either the trivial electronic insulator, or the topological insulator. This is achieved by a very simple perturbation: charge density modulations that only break translational symmetry. Interestingly, the $\mathbb{Z}_2$ topological index depends on the sign of the charge density perturbation. This transition, controlled by the sign of the mass perturbation, may be understood as a band inversion at an odd number of time-reversal invariant momenta, i.e., all the $M'$ points. In this respect, the Dirac semimetal state significantly differs from graphene, which has double Dirac nodes (counting spin) at each of the two $K$ points. The present spin-orbit coupled Dirac semimetal has a double node at the \emph{three} $M'$ points, i.e., three flavors of double nodes. In this respect it bears some similarity to the $(111)$ surface states of topological crystalline insulators in the SnTe material class~\cite{hsieh12,liu13,li16}, which are located at the $M$-points of the surface BZ. It will be interesting to further develop the Dirac theory of the three $M'$ points and consider the effect of various symmetry breaking perturbations. 

We have shown that the classical spin state analogs of non-coplanar $\beta$ phase spin density waves are symmetric spin states. Recently, the latter were shown to be the classical long-range ordered limits of time-reversal symmetry broken or chiral spin liquid phases~\cite{messio13}. Very recently, it was shown that quantum disordering the non-coplanar chiral spin state realized in a chiral spin model on the honeycomb lattice can result in a chiral spin liquid~\cite{hickey15}. Furthermore, recent theoretical work has considered quantum disordering the \emph{electronic} chiral spin density wave state of a quarter doped honeycomb lattice model, and found that the resulting spin-charge liquid state is topologically ordered~\cite{jiang14}. This establishes an exciting connection between the electronic $\beta$-phase spin density waves and spin liquid physics. In particular, it will be interesting to explore to connection of the scalar spin-current $d$-wave state, the time-reversal invariant analog of the chiral spin density wave, to the physics of spin(-charge) liquids. In this respect, we note that, since the $d$-wave state is time-reversal invariant, it can be converted into a triplet spin correlation function as~\cite{nayak00}
\begin{gather}
\order{\vec{S}(\vec{k}+\vec{Q}_\mu)\times \vec{S}(\vec{k})} =  \Delta \; \vec{d}(\vec{k}),
\end{gather}
where the $\vec{d}(\vec{k})$ vector lives in spin space and the components are given by Eq.~\eqref{eq:trispinf2}.
Therefore, the triplet density waves studied in this work give rise to intriguing questions to be addressed in future work.

\begin{acknowledgements}
I have benefited greatly from helpful and stimulating conversations and with L. Fu, C. Ortix, J. van Wezel, J. van den Brink, M. Daghofer, J. Hoon Han, Y. Ran, T. Iadecola and C. Chamon. I gratefully acknowledge K. Barros,  G.W. Chern and C.D. Batista for collaborations on related subjects. 
I am particularly indebted to C. Ortix and J. van den Brink for a careful reading of the manuscript and thoughtful comments. This work was supported by the Netherlands Organization for Scientific Research (NWO).
\end{acknowledgements}

\appendix


\section{$M$-point representation of hexagonal symmetry\label{app:malgebra}}


The $M$-point representation of hexagonal symmetry is specified by the action of elements of the symmetry group on the vector $\vet{v}=\vet{v}(\vet{x})$ defined as 
\begin{gather}
\vet{v}(\vet{x}) = \begin{pmatrix} \cos \vet{M}_1\cdot\vet{x} \\ \cos \vet{M}_2\cdot\vet{x}  \\  \cos \vet{M}_3\cdot\vet{x}  \end{pmatrix}.
\end{gather} 
The translations $T(\vet{a}_i)$, where $\vet{a}_i$ ($i=1,2,3$) are the elementary lattice vectors, are represented by the matrices $G_i$ defined through the equation
\begin{gather} \label{eq:defG}
\vet{v}(\vet{x}+\vet{x}_i) \equiv G_{i}\vet{v}(\vet{x}) , \quad i=1,2,3.
\end{gather}
Explicitly, $G_1$ and $G_2$ are given by
\begin{gather} 
G_{1}= \begin{pmatrix}-1 && \\ &-1& \\ &&1 \end{pmatrix}, \;  G_{2} = \begin{pmatrix}1 && \\ &-1& \\ &&-1 \end{pmatrix}.
\end{gather}
The matrices $G_i$ inherit the algebraic properties of the translations. They satisfy $G^2_i = 1$, they mutually commute and multiplication of two of them gives the third, i.e. $G_1G_2=G_3$.

The rotations and reflections can be expressed in terms of the generators $C_6$ and $\sigma_v$. The action of $C_6$ is defined as $\vet{v}'(\vet{x}) = \vet{v}(C^{-1}_6\vet{x})$ and is given by the matrix $X$,
\begin{gather} \label{eq:defX}
\vet{v}(C^{-1}_6\vet{x}) = X \vet{v}(\vet{x}), \quad  X = \begin{pmatrix}  0 & 1 & 0 \\ 0 & 0 & 1 \\1 & 0 & 0 \end{pmatrix}
\end{gather}
Note that $X$ has the property $X^{3}=1$ and thus $X^{-1} = X^2$. In addition, one has $X^{-1} =  X^T$, where $X^T$ is the transpose. It thus follows that $\vet{v}(C^{-1}_3\vet{x}) = X^2 \vet{v}(\vet{x}) =  X^T \vet{v}(\vet{x})$. For the reflection $\sigma_v$ we define the matrix $Y$ as
\begin{gather}  \label{eq:defY}
\vet{v}(\sigma^{-1}_v \vet{x}) = Y\vet{v}(\vet{x}), \quad  Y = \begin{pmatrix}  0 & 0 & 1 \\ 0 & 1 & 0 \\ 1 & 0 & 0 \end{pmatrix}.
\end{gather}
All rotations and reflections can be represented by a product of powers of $X$ and $Y$, i.e. $X^mY^n$. An arbitrary string of $X$ and $Y$ matrices can be brought into this form using $(XY)^2=1$, which is equivalent to $XY=YX^T$. 

The set of generator matrices $\{ G_1, X, Y\}$ (translations $G_2$ and $G_3$ can be written as products of the generators) defines an embedding of the group $C'''_{6v}$ in $O(3)$, the group of orthogonal matrices in three dimensions. Clearly, this mapping is not invertible, as the inversion $C_2 = C^3_6$ is mapped to identity through $X^3=1$. An invertible embedding is obtained by redefining the set of generators as $\{ G_1, -X, Y\}$, i.e., associating the $O(3)$ matrix $-X$ with $C_6$. In this way, two-fold rotation $C_2 \in C'''_{6v}$ is mapped to the inversion $P \in O(3)$. 

As we explained in the main text, the key property of such mapping is that it establishes an exact isomorphism between $C'''_{6v}$ and the cubic group $O_h$, a subgroup of $O(3)$. Indeed, direct inspection shows that the character tables of both groups are identical, which is a manifestation of the isomorphism. When mapped onto elements of the cubic group, the translations $T(\vet{a}_i)$ (represented by $G_i$) are given by two-fold rotations around the $x$, $y$ and $z$ axes. The six-fold rotations $C_6$ are interpreted as three-fold rotations about body diagonals of the cube combined with inversion (i.e., $-X$). The reflection $Y$ is mapped to a two-fold rotation about the axis $\hat{x}-\hat{z}$ combined with rotation. 
  
The mapping to the cubic group implies that the representations defined by $\{ G_1, -X, Y\}$ and $\{ G_1, X, Y\}$ can be labeled using the $O_h$ character table. The faithful representation generated by $\{ G_1, -X, Y\}$ is equal to the matrix representation of $(x,y,z)$ under cubic symmetry and hence given by $T_{1u}$. The representation generated by $\{ G_1, X, Y\}$ corresponds to the cubic representation $T_{2g}$, which describes the symmetry of the $d$-orbitals $(yz,zx,xy)$. In general, the 3D representations of the cubic group, $\{ T_{1g}, T_{2g}, T_{1u}, T_{2u} \}$, are in correspondence with the 3D representations of $C'''_{6v}$, $\{ F_{2}, F_{1}, F_{3}, F_{4} \}$ (in that order). The latter describe translational symmetry broken $M$-point modulations. The mapping between cubic and hexagonal symmetries is summarized in Table~\ref{tab:mapping}. 

\begin{table}[t]
\centering
\begin{ruledtabular}
\begin{tabular}{ccc}
 & Hexagonal $C'''_{6v}$ &  Cubic $ O_h$   \\ 
\hline
Reps. & $F_1$, $F_2$, $F_3$, $F_4$  & $T_{2g}$, $T_{1g}$, $T_{1u}$, $T_{2u}$   \\
$G_1$, $G_2$, $G_3$ & Translations $T(\vec{a}_i)$ & Twofold rotations $C_{2i}$  \\
$-I_3$ & Twofold Rotation $C_2$ & Inversion $P$  \\
$X$, $X^T$  & Twofold rotations & Threefold rotations \\
&   (Principal $C_3$) & (Body diagonal $C_3$)  \\
$Y$   & Reflection $\sigma_v$  & Roto-reflection $PC'_2$ \\
\end{tabular}
\end{ruledtabular}
 \caption{Table summarizing the mapping between the hexagonal group $C'''_{6v}$ and the cubic group $O_h$. Here $I_3$ is the identity matrix. }
\label{tab:mapping}
\end{table}

An equivalent definition of the $M$-point representation follows from considering the action of symmetry operations on the van Hove electron operator $\hat{\Phi}  $ introduced in the main text [see Eq.~\eqref{eq:vanhovehexa}] and given by  
\begin{gather} 
\hat{\Phi} =  \begin{pmatrix} \anni{\sigma }{\vet{M}_1} \\ \anni{\sigma }{\vet{M}_2}   \\ \anni{\sigma }{\vet{M}_3}   \end{pmatrix}.
\end{gather}
Evaluating the action of the generators of $C'''_{6v}$ on the $M$-point index $\mu$, i.e. $\vet{M}_\mu$ one finds
\begin{align} 
T(\vet{x}_1) \; : \;  \hat{\Phi} &\rightarrow \; G_1 \hat{\Phi}, \nonumber \\
C_6 \; : \;  \hat{\Phi} &\rightarrow \; X \hat{\Phi} \nonumber \\
\sigma_v \; : \;  \hat{\Phi} &\rightarrow \; Y \hat{\Phi},  \label{eq:vanhoveapp}
\end{align}
where $G_1$, $X$ and $Y$ are the matrices given in~\eqref{eq:defG}--\eqref{eq:defY}. We stress that here we only consider the action of symmetries on the $M$-point index, and for the moment disregard the action on the internal spin degree of freedom. 

Starting from $\hat{\Phi}$, the mapping to cubic symmetry can be formulated in terms of a $d$-orbital electron operator $\hat{\Upsilon}$ defined as, 
\begin{gather} 
\hat{\Upsilon} =  \begin{pmatrix} \hat{\psi}_{yz\sigma }\\ \hat{\psi}_{zx\sigma }  \\ \hat{\psi}_{xy\sigma }  \end{pmatrix}, 
\end{gather}
The $M$-point representation generated by $\{ G_1, X, Y\}$ arises from considering the action of symmetries on $\hat{\Phi}$, and as a result of the isomorphism between $C'''_{6v}$ and $O_h$, it arises equivalently from considering the action of $O_h$ elements on $\hat{\Upsilon}$. Specifically, one finds that
\begin{align} 
C_2 \; : \;  \hat{\Upsilon}  &\rightarrow \; G_1 \hat{\Upsilon} , \nonumber \\
PC_3 \; : \;  \hat{\Upsilon}  &\rightarrow \; X \hat{\Upsilon} , \nonumber \\
PC'_2 \; : \;  \hat{\Upsilon}  &\rightarrow \; Y \hat{\Upsilon} ,  \label{eq:cubicsym}
\end{align}
where $C_2$ is a two-fold rotation about the principal $z$-axis, $P$ is the inversion, $C_3$ is a three-fold rotation about a body diagonal, and $C'_2$ is another (inequivalent) two-fold rotation. We conclude that $\hat{\Upsilon}$ transforms in exactly the same way under $O_h$ symmetry as $\hat{\Phi}$ under $C'''_{6v}$ symmetry. Note that the representation matrices act on the orbital degree of freedom and not on spin, and the equivalence of $\hat{\Upsilon}$ and $\hat{Phi}$ pertains to the spatial (i.e., orbital) degree of freedom. 

Insofar as spin is concerned, $\hat{\Upsilon}$ transforms according to the cubic double group. Specifically, each element $g \in O_h$ is associated with $U_g \in SU(2)$ such that $\hat{\Upsilon}  \rightarrow \; U_g O_g \hat{\Upsilon} $, where $O_g$ the matrix representation of $g$ obtained from $\{ G_1, X, Y\}$. For instance, the three two-fold rotations about the principal axis have $\{ U_{C^{(x)}_2},U_{C^{(y)}_2},U_{C^{(z)}_2} \} = \{ -i\sigma^1 ,-i\sigma^2 ,-i\sigma^3 \} $. In addition, the rotation $X$ is accompanied with $U_X = e^{i\pi\sigma^3/4} e^{i\pi\sigma^2/4}$, and the rotation $-Y$ with $U_Y= e^{i\pi\sigma^2/4}e^{-i\pi\sigma^3/2}=-i e^{i\pi\sigma^2/4}\sigma^3$.

We conclude this appendix by providing explicit expression for the matrices of fermion bilinears $\hat{\Lambda}$ given by $\hat{\Lambda}= \hat{\Phi}^\dagger_\mu \Lambda_{\mu\nu}  \hat{\Phi}_\nu$. Here $\Lambda$ is an Hermitian matrix and the space of these $M$-point Hermitian matrices is spanned by the Gell-Mann matrices, the generators of SU(3). In this work we choose to group them in three sets defined by $\vet{\Lambda}_a$, $\vet{\Lambda}_b$ and $\vet{\Lambda}_c$. They are given by
\begin{gather}
\Lambda^1_{a} =  \begin{pmatrix} 0 & 1 & 0 \\ 1 & 0& 0 \\ 0&0&0 \end{pmatrix}, \; \Lambda^2_{a} = \begin{pmatrix} 0 & 0 & 0 \\ 0 & 0& 1 \\ 0 & 1&0 \end{pmatrix}, \; \Lambda^3_{a} = \begin{pmatrix} 0 & 0 & 1 \\ 0 & 0& 0 \\ 1 & 0&0 \end{pmatrix}  \nonumber \\
\Lambda^1_{b} =  \begin{pmatrix} 0 & -i & 0 \\ i & 0& 0 \\ 0&0&0 \end{pmatrix}, \; \Lambda^2_{b} = \begin{pmatrix} 0 & 0 & 0 \\ 0 & 0& -i \\ 0 & i&0 \end{pmatrix} , \; \Lambda^3_{b} = \begin{pmatrix} 0 & 0 & i \\ 0 & 0& 0 \\ - i & 0&0 \end{pmatrix}  \nonumber \\
\Lambda^1_{c} =\begin{pmatrix} 1 & 0 & 0 \\ 0 & -1 & 0 \\ 0&0&0 \end{pmatrix}, \; \Lambda^2_{c} =   \frac{1}{\sqrt{3}} \begin{pmatrix} 1 & 0 & 0 \\ 0 & 1 & 0 \\ 0 & 0& -2 \end{pmatrix}   . \label{eq:gellmann}
\end{gather}
With the help of the symmetry transformation properties of Eq.~\eqref{eq:vanhoveapp}, it is straightforward to establish that $\vet{\Lambda}_a$ transforms as $F_1$ and $\vet{\Lambda}_b$ as $F_2$.


\section{Low-energy Dirac theory at the $M'$-points\label{app:dirac}}

\subsection{Proof of degeneracy}

The proof of the symmetry protected denegeracy at the $\vet{M}'_2$ point of the folded BZ requires evaluating the effect of symmetries leaving $\vet{M}'_2$ invariant on the electron operator of Eq.~\eqref{eq:m2operator}. These symmetries are inversion $C_2$, reflection $\sigma_v$ and translations $T(\vet{a}_i)$. The action of $C_2$ is given by
\begin{gather}
C_2 \quad \rightarrow \quad  \begin{pmatrix} \anni{\sigma}{-\vet{M}'_2} \\  \anni{\sigma}{-\vet{M}'_2+\vet{M}_1} \\   \anni{\sigma}{-\vet{M}'_2+\vet{M}_2} \\  \anni{\sigma}{-\vet{M}'_3+\vet{M}_3}    \end{pmatrix} = \nu^1 \hat{\Phi}_{M'}.
\end{gather}
From this we conclude that the Hamiltonian at $\vet{M}'_2$ can only have terms $\sigma^i\tau^j$ or $\sigma^i\tau^j\nu^1$, where it is understood that $i,j=0,1,2,3$. From Appendix~\ref{app:malgebra} we know that the translation $T(\vet{a}_2)$ is associated with $G_2$, i.e., a rotation around the $x$ axis by $\pi$, and as a result the symmetry $T(\vet{a}_2)$ acts as [disregarding U(1) phases]
\begin{gather}
T(\vet{a}_2) \quad \rightarrow \quad  \sigma^1 \nu^3 \hat{\Phi}_{M'}.
\end{gather}
This leaves us with the allowed terms $\tau^j$, $\sigma^1\tau^j$, $\sigma^2\tau^j\nu^1$ and $\sigma^3\tau^j\nu^1$. Similarly, the translation $T(\vet{a}_3)$ is associated with $G_3$ and therefore $e^{-i\pi\sigma^2/2}$. Hence, the action of the translation symmetry is
\begin{gather}
T(\vet{a}_3) \quad \rightarrow \quad   \sigma^2 \tau^3 \hat{\Phi}_{M'},
\end{gather}
which leaves us with the following allowed terms
\begin{gather*}
\tau^3,\; \sigma^1\tau^1,\; \sigma^1\tau^2,\; \sigma^2\nu^1,\;  \sigma^2\tau^3\nu^1, \\
\sigma^3\tau^1\nu^1,\; \sigma^3\tau^2\nu^1.
\end{gather*}
We are left with two reflections leaving $\vet{M}'_2$ invariant. We consider $C_2\sigma_v$, the action of which on $\hat{\Phi}_{M'}$ is 
\begin{gather}
C_2\sigma_v \quad \rightarrow \quad   e^{i\pi \sigma^2/4}\sigma^3 \begin{pmatrix} 1 &0&0&0 \\0 &0&0&1  \\ 0&0&1 &0 \\ 0&1&0&0   \end{pmatrix} \hat{\Phi}_{M'}.
\end{gather}
The spin rotation $U_{Y} = -i e^{i\pi \sigma^2/4}\sigma^3 $ is the global spin rotation associated with $Y$ (see also Appendix~\ref{app:malgebra}). This transformation property immediately leads to the exclusion of $\sigma^2\tau^3\nu^1$ and $\sigma^2\nu^1$. The term $\tau^3$ is clearly left invariant. The remaining four terms must be combined in order to represent invariant terms, and in the end we find the following three terms allowed by symmetry
\begin{gather*}
\tau^3,\; \tau^1(\sigma^1 - \sigma^3\nu^1),\; \tau^2(\sigma^1 - \sigma^3\nu^1).
\end{gather*}
Applying a basis transformation $e^{-i\pi\sigma^1\nu^1/4}e^{i\pi\sigma^3/8}$ brings them into the form $\sigma^1\tau^1$, $\sigma^1\tau^2$ and $\tau^3$. Clearly, these three matrices mutually anti-commute and as result any linear combination of these terms, i.e., the most general Hamiltonian allowed by spatial symmetry, can only have two eigenvalues $\epsilon$ and $-\epsilon$. Each eigenvalue must be fourfold degenerate, proving the symmetry protection of the double Dirac nodes at $\vet{M}_2$. Clearly, the same is true for the other $M'$ points. 

We can exclude one more term using time-reversal symmetry $\mathcal{T}$. Time-reversal acts as
\begin{gather}
\mathcal{T} \quad \rightarrow \quad   -i\sigma^2 \nu^1 \hat{\Phi}_{M'},
\end{gather}
from which we conclude that the only allowed terms are $\tau^3 $ and $\tau^2(\sigma^1 - \sigma^3\nu^1)$.

\subsection{Low-energy Dirac theory}

The low-energy theory of the mean-field Dirac nodes at $\vet{M}'_2$ is constructed by expanding the mean-field band structure to linear order around $\vet{M}'_2$. The first step is to diagonalize the Hamiltonian $\mathcal{H}(\vet{M}'_2)$, which is given by $\mathcal{H}(\vet{M}'_2)=-2t \tau^3 + \Delta(- \sigma^1\tau^2 + \sigma^3\tau^2\nu^1)$.
We perform a basis transformation $U^\dagger \mathcal{H}(\vet{M}'_2) U$ with $U = e^{i\pi \sigma^1\nu^1/4} e^{i\pi \sigma^3/8} e^{-i\pi \sigma^1/4}$. This yields the Hamiltonian
\begin{gather}
\mathcal{H}(\vet{M}'_2)=-2t \tau^3 + \sqrt{2}\Delta \tau^2\sigma^3  \equiv - \xi \tau^3 + \eta \tau^2\sigma^3 
\end{gather}
As proven earlier, the Hamiltonian has two eigenvalues, $\epsilon_{\pm} = \pm \epsilon \equiv \pm \sqrt{\xi^2 + \eta^2}$. The eigenvectors corresponding to $\epsilon_+$, i.e. the subspace of the relevant Dirac nodes, are given by $\ket{\varphi_{mn} }$, 
\begin{eqnarray}
\ket{\varphi_{11} }  & =& (u,0,-iv,0,0,0,0,0 ), \nonumber \\
\ket{\varphi_{12} }  & =& (0,0,0,0,u,0,-iv,0 ),  \nonumber \\
\ket{\varphi_{21} }  & =&  (0,u,0,iv,0,0,0,0 ),  \nonumber \\
\ket{\varphi_{22} }  & =&  (0,0,0,0,0,u,0,iv ) ,
\end{eqnarray}
where $u$ and $v$ are defined as 
\begin{gather}
u = \frac{1}{\sqrt{2}}\sqrt{1-\frac{\xi}{\epsilon}}, \quad v = \frac{1}{\sqrt{2}}\sqrt{1+\frac{\xi}{\epsilon}}.
\end{gather}
Note that if $\Delta \rightarrow 0$ (i.e., $\eta \rightarrow 0$) one has $v=1$ and $u=0$, as expected. 
 
The next step is to expand the mean-field Hamiltonian $\mathcal{H}(\vet{k})$ in small $\vet{q}$ with respect to $\vet{M}'_2$, retaining only the linear terms. We then perform the same basis transformation $U$ and project the expanded Hamiltonian into the subspace given by $\ket{\varphi_{mn} }$. The $q$-linear part of the Hamiltonian at $\vet{M}'_2$ is
\begin{align}
\mathcal{H}_q =& \xi (q_2\nu^3-q_1\nu^3\tau^3 ) + \eta(q_1 \nu^3\tau^2\sigma^1-q_1\nu^2\tau^3\sigma^2 )/\sqrt{2}\nonumber \\
&+ \eta(q_2\nu^2\tau^1\sigma^3-q_2\nu^2\sigma^2) /\sqrt{2}. 
\end{align}
To express the resulting Dirac Hamiltonian in terms of effective valley and pseudospin degrees of freedom, we define two sets of Pauli matrices, $\tilde{\nu}$ and $\tilde{\tau}$, which act on $m$ and $n$ of $\ket{\varphi_{mn} }$, respectively. In addition, we take $q_\pm  = q_1 \pm q_2$, where $q_i=\vet{q}\cdot \vet{a}_i$. We find the Hamiltonian
\begin{gather}
\mathcal{H}(\vet{q}) = (v_1 q_- +v'q_+)\tilde{\nu}^1\tilde{\tau}^2+(v_2 q_+ -v'q_-)\tilde{\nu}^3\tilde{\tau}^2
\end{gather}
with $v_1 = \xi u^2 + \eta v^2 /\sqrt{2}$, $v_2 = \xi v^2 + \eta u^2 /\sqrt{2}$, and $v' = 2\eta uv /\sqrt{2}$. As a result, in the limit $\Delta \rightarrow 0$, i.e., the absence of any order, $v_2=\xi$ and $v_1,v'=0$. This is as expected, since $q_-$ is in the direction of the undistorted Fermi surface, implying there is no dispersion in that direction. 

The inversion $C_2$ is given by $\nu^1 \hat{\Phi}_{M'}$. Projecting $\nu^1$ into the Dirac spinor subspace defined by $\ket{\varphi_{mn} }$ we find $\tilde{\tau}^1$. Projecting the perturbation $\delta\mathcal{H}$ given in Eq.~\eqref{eq:perturb} into the same subspace yields $m \tilde{\tau}^1$. This is consistent with the fact that $\delta\mathcal{H}$ is symmetric under inversion. More importantly, this demonstrates that $m$ controls what the inversion eigenvalue of the occupied Kramer's doublet is. 

An alternative way to understand the dependence of the topological invariant (in this case given by the Fu-Kane formula~\cite{fu07}) on $\text{sgn}(m)$ is to start from the charge density modulations $\delta\mathcal{H}$ and consider the $M$-point electrons at $\Gamma$:
\begin{gather}
 \hat{\Phi}_{\Gamma} = \begin{pmatrix} \anni{\sigma }{\vet{M}_1} \\ \anni{\sigma }{\vet{M}_2}   \\ \anni{\sigma }{\vet{M}_3}   \end{pmatrix} \equiv \begin{pmatrix}\hat{\psi}_{1\sigma } \\ \hat{\psi}_{2\sigma }    \\ \hat{\psi}_{3\sigma }   \end{pmatrix}.
\end{gather}
In the presence of the perturbation $\delta\mathcal{H}$ (assuming $\Delta=0$), the $\hat{\Phi}_{\Gamma} $ states split into non-degenerate level and a degenerate doublet $(\hat{\Psi}_{1\sigma},\hat{\Psi}_{2\sigma})$~\cite{venderbos15}, the latter given by
\begin{eqnarray}
 \hat{\Psi}_{1\sigma} &=&  ( \hat{\psi}_{1\sigma} + \hat{\psi}_{2\sigma} -2 \hat{\psi}_{3\sigma}  )/\sqrt{6} , \nonumber \\ 
 \hat{\Psi}_{2\sigma} &=&  ( - \hat{\psi}_{1\sigma} + \hat{\psi}_{2\sigma} ) /\sqrt{2}.
\end{eqnarray}
The sign of $m$ determines the relative energetic ordering of the doublet and the non-degenerate level. 

If $m<0$, the doublet is higher in energy, and the Fermi level is at the semimetallic quadratic band crossing point defined by $(\hat{\Psi}_1,\hat{\Psi}_2)$ and governed by the quadratic band crossing Hamiltonian $\mathcal{H}(\vet{q}) \sim (q_x^2 - q^2_y )\tilde{\tau}^3 + 2q_xq_y\tilde{\tau}^1 $ (see Ref.~\onlinecite{venderbos15}). Here, $\tilde{\tau}^3=\pm 1$ labels the states $\hat{\Psi}_{1,2}$. 

We can now consider finite $\Delta$, i.e., finite $\Delta_{A_1}$ in Eq.~\eqref{eq:trispinf2}. Projecting the ``perturbation'' coming from $\Delta$ into the subspace given by $(\hat{\Psi}_{1\sigma},\hat{\Psi}_{2\sigma})$ we find $\Delta \vet{n}\cdot \vet{\sigma}\tilde{\tau}^2$, with $\vet{n} = (1,1,1)$. This is recognized as a QSH gap of a quadratic band crossing: $\tilde{\tau}^2$ constitutes the quantum anomalous Hall gap, and $\hat{n}\cdot \vet{\sigma}$ gives it opposite sign for the two spin projections. This proves that the resulting state, which is adiabatically connected to the Dirac semimetal with gap $m<0$ at the $M'$ points, is a topological insulator state.

In contrast, had we assumed $m>0$, the quadratic band crossing point defined by $(\hat{\Psi}_1,\hat{\Psi}_2)$ would be fully occupied (i.e., the Fermi level would \emph{not} be precisely at the quadratic band crossing point) and the insulating state with finite $\Delta$ would be adiabatically connected to the trivial insulator defined by $\delta\mathcal{H}$ with $m>0$.

\section{Derivation of symmetric classical spin states\label{app:classical}}

We briefly review the derivation of charge order representations,
introduced in Ref.~\onlinecite{venderbos15}, based
on the example of the honeycomb and kagome lattices discussed in
Section~\ref{sec:classical}. 

Modulations with $M$-point propagation vectors lead to a quadrupling of the unit
cell. In case of the honeycomb lattice the enlarged unit cell contains
$n_s= 4 \times 2  =8$ sites, whereas in case of the kagome lattice the enlarged
unit cell contains $n_s= 4 \times 3  =12 $. We label all sites and collect them in a
vector $\vet{s}$ given by $\{ s_i \}^{n_s}_{i=1} $. Extended point
group operations will permute the sites and the permutation matrices define a
representation of extended point group. We write the representation as
$\mathcal{P}^M_{s}$ and it has dimension $n_s$. The superscript $M $
is meant to indicate $M$-point modulation (unit cell quadrupling). 

The representation $\mathcal{P}^M_{s}$ is reducible and can be
decomposed into irreducible representations of the extended point
group. In case of the honeycomb one finds
\begin{gather}  \label{eq:hexasitem}
\mathcal{P}^{M}_{s} = A_1 + B_2 + F_1 + F_4,
\end{gather}
whereas the kagome lattice yields
\begin{gather}   \label{eq:kagsitem} 
\mathcal{P}^{M}_{s} = A_1 + E_2 + F_1 + F_3 + F_4.
\end{gather}
The $F_i$ signal translational symmetry breaking and constitute the
$M$-point modulated content of the decomposition. These
representations are listed in Table~\ref{tab:classical}. We observe
that the honeycomb lattice has two independent representations $F_1$
and $F_4$, and the kagome lattice admits three, $ F_1 $, $ F_3$, and $
F_4$. From this we conclude that the former admits two classical spin
liquid states, whereas the latter admits three.

\section{Extended point groups and character tables\label{app:gt}}

Here, we provide a basic review of the essentials of extended point group symmetry used in the main text. The crystal point group of 2D hexagonal materials is $C_{6v}$, which is identical to the dihedral group $D_6$ for spinless electrons. For spinful electrons the point groups $D_6$ and $C_{6v}$ are technically distinct, however, for the purpose of this work we consider them equivalent, as our results are independent of technical differences, and focus on the point group $C_{6v}$. Note that time reversal acts as $\mathcal{T} = i \sigma^2 \mathcal{K}$ ($\mathcal{K}$ is complex conjugation). 

The group $C_{6v}$ is generated by a six-fold rotation $C_6$ and a reflection $\sigma_v$, where the reflection is defined as $(x,y)\rightarrow (x,-y)$. The space group $S$ is given by all point group elements and all translations over lattice vectors $\vet{x} $, i.e.,  $T(\vet{x})$, where the lattice vectors are generated by $\vet{a}_1$ and $\vet{a}_2$. As a result, the translation subgroup $T$ is generated by $T(\vet{a}_1)$ and $T(\vet{a}_2)$. 

In general, a point group $G$ can be obtained as the factor group of the space group $S$ and the translation subgroup $T$. The \emph{extended} point groups are obtained as the factor group of the space group and a modified translation subgroup $\tilde{T}$, defined as the group of translations compatible with ordering vectors, or equivalently, with the enlarged unit cell. All translation in $\tilde{T}$ map the enlarged unit cell to itself. The enlargement is fixed by the ordering vectors, in our case the $M$-point vectors. Hence, the extended point group is defined as $\tilde{G} = S/ \tilde{T}$. Clearly, $\tilde{G} $ is larger than $G$, as the translations in $T$ but no longer in $\tilde{T}$ are now in $\tilde{G}$.

In this work, we consider hexagonal symmetry, $C_{6v}$, and ordering at the $M$ points. The latter implies translational symmetry breaking such that $\widetilde{T}$ is generated by $T(2\vet{a}_1)$ and $T(2\vet{a}_2)$. The translations $T(\vet{a}_1) \equiv t_1$, $T(\vet{a}_2) \equiv t_2$ and $T(\vet{a}_1+\vet{a}_2) \equiv t_3$ are added to the point group. Since three translations are added to the point group, we denote the extended point group of $C_{6v}$ as $C'''_{6v}$. The character table of $C'''_{6v}$ is given in Table~\ref{tab:cppp6v}.

In the main text, we use an isomorphism between the hexagonal extended point group $C'''_{6v}$ and the cubic point group $O_h$. This connection is discussed in more detail in Appendix~\ref{app:malgebra}. The character table of the cubic group is given in Table~\ref{tab:charactab_Oh}.

\subsection{Lattice angular momentum basis functions\label{app:basis}}

\begin{table}[t]
\centering
\begin{ruledtabular}
\begin{tabular}{cccc}
Rep. & Type & Label  & Expression \\ 
\hline
$A_1$  & $s$ & $\lambda_s(\vet{k})$ & $(\cos k_1+\cos k_2 + \cos k_3)/\sqrt{3}$ \\
$E_2$   & $d_{x^2-y^2}$ &$\lambda_{d_1}(\vet{k})$& $(\cos k_1+\cos k_2 -2\cos k_3 )/\sqrt{6}$  \\
 & $d_{xy}$ & $\lambda_{d_2}(\vet{k})$  & $(\cos k_1-\cos k_2 )/\sqrt{2}$  \\
\end{tabular}
\end{ruledtabular}
 \caption{Lattice angular momentum form factors transforming as representations of $C_{6v}$, corresponding to nearest neighbor hopping on the triangular lattice. We defined $k_i = \vet{k}\cdot \vet{a}_i$ with $\vet{a}_i$ given in Sec.~\ref{sec:tripletm}. Note that $(\lambda_{d_1},\lambda_{d_2}) \sim (k_x^2-k_y^2,2k_xk_y)$ when expanded in $\vet{k}$. }
\label{tab:trifunctions}
\end{table}


The triangular lattice angular momentum form factor functions, used to express condensate functions, are given in Table~\ref{tab:trifunctions}. Table~\ref{tab:trifunctions} lists the functions that transform as representations of $C_{6v}$ and correspond to form factors originating from triangular lattice nearest-neighbor (e.g., $\vet{a}_i$) hopping. 

%
%

\subsection{Character tables\label{app:chartab}}

For completeness and convenience, here we reproduce the character tables of the extended point groups $C'''_{6v}$ (see Table \ref{tab:cppp6v}) and $O_h$ (see Table \ref{tab:charactab_Oh}).

\begin{table*}[t]
\centering
\begin{ruledtabular}
\begin{tabular}{c|rrrrrrrrrr}
 Conjugacy class         &   \multicolumn{1}{c}{ $\mathcal{C}'''_1$ }   &  \multicolumn{1}{c}{ $\mathcal{C}'''_2$ }   &  \multicolumn{1}{c}{ $\mathcal{C}'''_3$ } & \multicolumn{1}{c}{ $\mathcal{C}'''_4$ }   & \multicolumn{1}{c}{ $\mathcal{C}'''_5$ } & \multicolumn{1}{c}{ $\mathcal{C}'''_6$ }  & \multicolumn{1}{c}{ $\mathcal{C}'''_7$ }  &\multicolumn{1}{c}{ $\mathcal{C}'''_8$ }  & \multicolumn{1}{c}{ $\mathcal{C}'''_9$ }  & \multicolumn{1}{c}{ $\mathcal{C}'''_{10}$ }  \\ 
\cline{2-11}
      Point group         &      &  $t_1$, $t_2$ & &  $t_1C_2$, $t_2 C_2$   & $t_i C_3$, $t_iC^{-1}_3$ & $t_i C_6$, $t_iC^{-1}_6$ & $ 3\sigma_v$, $t_1\sigma_{v2}$ & $ t_1\sigma_{v}$, $ t_2\sigma_{v}$  &$3\sigma_{d}$, $t_2\sigma_{d1}$ &  $t_1\sigma_{d1}$, $t_3\sigma_{d1}$  \\
  $C'''_{6v} $   & $ I $& $t_3$ & $C_2$  &$t_3 C_2$  & $C_3$, $C^{-1}_3$ & $C_6$, $C^{-1}_6$  & 
 $t_2 \sigma_{v3} $,  $t_3 \sigma_{v1}$ &  $t_2 \sigma_{v2} $,  $t_3 \sigma_{v2}$ &   $t_3 \sigma_{d2}$,  $t_1\sigma_{v3}$ &   $t_1\sigma_{d2}$, $t_2\sigma_{d2}$   \\ 
&  &  &   &  &  &  & 
 &  $t_1 \sigma_{v3} $,  $t_3 \sigma_{v3}$ &    &   $t_2\sigma_{d3}$, $t_3\sigma_{d3}$   \\ [1ex]
\hline 
 $A_1$ & $1$& $1$ & $1$ & $1$  & $1$ & $1$  &$1$   &$1$ & $1$ &  $1$ \\ 
 $A_2$ & $1$& $1$ & $1$ & $1$  & $1$ & $1$  &$- 1$ & $-1$ & $-1$ & $-1$ \\ 
 $B_1$ & $1$& $1$ & $-1$& $-1$ & $1$ & $-1$ &$ 1$  & $1$ & $-1$ & $-1$  \\
 $B_2$ & $1$& $1$ & $-1$& $-1$ & $1$ & $-1$ &$- 1$ & $-1$ & $1$ & $1$ \\
 $E_1$ & $2$& $2$ & $-2$& $-2$ & $-1$& $1$  &$ 0$  & $0$ & $0$ &  $0$ \\
 $E_2$ & $2$& $2$ & $2$ & $2$  & $-1$& $-1$ &$ 0$  & $0$ & $0$ &  $0$ \\
\hline
 $F_1$ & $3$& $-1$& $3$ & $-1$ & $0$& $0$  &$ 1$  & $-1$ & $1$ & $-1$  \\
 $F_2$ & $3$& $-1$& $3$ & $-1$ & $0$& $0$  &$ -1$ & $1$ & $-1$ &  $ 1$ \\
 $F_3$ & $3$& $-1$& $-3$& $1$  & $0$ & $0$  &$ 1$  & $-1$ & $-1$ &  $1$  \\
 $F_4$ & $3$& $-1$& $-3$& $1$  & $0$ & $0$  &$ -1$ & $1$ & $1$ & $-1$
\end{tabular}
\end{ruledtabular}
 \caption{Character table of the point group $C'''_{6v}$~\cite{hermele08}. Translations $t_1$ and $t_2$ correspond to $T(\vet{a}_1)$ and $T(\vet{a}_2)$, respectively. $t_3 = T(\vet{a}_1+\vet{a}_2)$. The irreducible representations that arise as a consequence of the added translations are $F_1$, $F_2$, $F_3$ and $F_4$, all three-dimensional.}
\label{tab:cppp6v}
\end{table*}

\begin{table*}[t]
\centering
\begin{ruledtabular}
\begin{tabular}{cc|rrrrrrrrrr}
 \multicolumn{2}{c|}{ Point group $O_h $} & $ I $&
 $3C^2_4$ & $6C_4 $ & $6C'_2 $ & $8C_3$ & $ P $& $3PC^2_4$ & $6PC_4 $ & $6PC'_2 $ & $8PC_3$ \\  
 Koster & Mulliken &  &  & & & & & & & & \\  [1ex]
\hline 
$\Gamma^+_1$ & $A_{1g}$ & $1$ & $1$  & $1$  &$1$    & $1$    & $1$ & $1$& $1$& $1$&$1$ \\ 
$\Gamma^+_2$ & $A_{2g}$   & $1$ & $1$  & $-1$  & $-1$   & $1$  & $1$ & $1$& $-1$& $-1$& $1$ \\ 
$\Gamma^+_3$ & $E_{g}$ & $2$ & $2$& $0$  & $0$ & $-1$  & $2$  & $2$& $0$& $0$& $-1$ \\ 
$\Gamma^+_4$ & $T_{1g}$  & $3$ & $-1$& $1$& $-1$   & $0$    &$3$ & $-1$& $1$& $-1$& $0$ \\
$\Gamma^+_5 $ & $T_{2g}$ & $3$ & $-1$  & $-1$& $1$ & $0$    &$3$ & $-1$& $-1$& $1$& $0$\\
\hline
$\Gamma^-_1$ & $A_{1u}$    & $1$ & $1$  & $1$  &$1$    & $1$    & $-1$ & $-1$& $-1$& $-1$&$-1$ \\ 
$\Gamma^-_2$ & $A_{2u}$    & $1$ & $1$  & $-1$  & $-1$   & $1$ & $-1$ & $-1$& $1$& $1$& $-1$ \\ 
$\Gamma^-_3$ &$E_{u}$      & $2$ & $2$& $0$  & $0$ & $-1$ & $-2$  & $-2$& $0$& $0$& $1$ \\ 
$\Gamma^-_4$ & $T_{1u}$    & $3$ & $-1$& $1$& $-1$   & $0 $   &$-3$ & $1$& $-1$& $1$& $0$ \\
$\Gamma^-_5$ & $T_{2u}$    & $3$ & $-1$  & $-1$& $1$ & $0$    &$-3$ & $1$& $1$& $-1$& $0$ \\
\end{tabular}
\end{ruledtabular}
\label{tab:charactab_Oh}
 \caption{Character table of the point group $O_{h}$. }
\end{table*}


\begin{thebibliography}{10}
\bibitem{note3} Examples of when it is not possible to "transfer" a translation to spin space are the uniaxial spin density wave or $\alpha$-phases (to be considered later in this section). These manifestly break translational symmetry.
\bibitem{note1} Note that this is in the triple-$M$ eigenbasis and not in the basis $\hat{\Phi}$.
\bibitem{note2}  Due to uniaxial spin order, time-reversal symmetry, which inverts all spin, can be compensated by a global rotation. This is equivalent to defining time-reversal simply as complex conjugation in each spin sector.
\end{thebibliography}
\end{document}